\documentclass[apjl,numberedappendix]{emulateapj}
\slugcomment{{\sc Accepted to ApJ:} November 19, 2016} 
\usepackage{multirow}
\usepackage[breaklinks,colorlinks,citecolor=blue]{hyperref}
\usepackage[all]{hypcap}

\usepackage{textcomp}

\shorttitle{Externally heated protostellar cores in Ophiuchus}
\shortauthors{Lindberg et al.}

\newcommand\nobr{\mbox{-}}
\newcommand\ctht{\textit{c}\nobr C$_3$H$_2$}
\newcommand\moo{~\hbox{\textmu}m}
\begin{document}


\title{Externally heated protostellar cores in the Ophiuchus star-forming region\footnotemark[*]}\footnotetext[*]{Based on observations with the Atacama Pathfinder EXperiment (APEX) telescope. APEX is a collaboration between the Max Planck Institute for Radio Astronomy, the European Southern Observatory, and the Onsala Space Observatory.}


\author{
Johan E. Lindberg\altaffilmark{1},
Steven B. Charnley\altaffilmark{1},
Jes K. J{\o}rgensen\altaffilmark{2},
Martin A. Cordiner\altaffilmark{1,3},
Per Bjerkeli\altaffilmark{2,4}
}

\affil{$^1$NASA Goddard Space Flight Center, Astrochemistry Laboratory, Mail Code 691, 8800 Greenbelt Road, Greenbelt, MD 20771, USA, johan.lindberg@nasa.gov}
\affil{$^2$Centre for Star and Planet Formation, Niels Bohr Institute and Natural History Museum of Denmark,\\ University of Copenhagen, {\O}ster~Voldgade 5-7, DK-1350 Copenhagen K, Denmark}
\affil{$^3$Department of Physics, Catholic University of America, Washington, DC 20064, USA}
\affil{$^4$Department of Earth and Space Sciences, Chalmers University of Technology, Onsala Space Observatory, 439 92 Onsala, Sweden}



\begin{abstract}
We present APEX 218~GHz observations of molecular emission in a complete sample of embedded protostars in the Ophiuchus star-forming region. To study the physical properties of the cores, we calculate H$_2$CO and \ctht\ rotational temperatures, both of which are good tracers of the kinetic temperature of the molecular gas. We find that the H$_2$CO temperatures range between 16~K and 124~K, with the highest H$_2$CO temperatures toward the \textit{hot corino} source IRAS~16293\nobr2422 (69--124~K) and the sources in the $\rho$~Oph~A cloud (23--49~K) located close to the luminous Herbig~Be star S~1, which externally irradiates the $\rho$~Oph~A cores. On the other hand, the \ctht\ rotational temperature is consistently low (7--17~K) in all sources. Our results indicate that the \ctht\ emission is primarily tracing more shielded parts of the envelope whereas the H$_2$CO emission (at the angular scale of the APEX beam; 3600~au in Ophiuchus) mainly traces the outer irradiated envelopes, apart from in IRAS~16293-2422, where the \textit{hot corino} emission dominates. In some sources, a secondary velocity component is also seen, possibly tracing the molecular outflow.

\end{abstract}





\section{Introduction}

Low-mass protostars form from the collapse of dense molecular clouds. In their youngest stages, such protostars are deeply embedded in an envelope of dust and gas. Investigations of molecular emission lines of this envelope can be used to study both chemical and physical characteristics of the protostar. The temperature of the large-scale envelope is generally low ($\sim10$--15~K; e.g. \citealt{bergin07}), but parts of the envelope can be heated by irradiation from the protostar itself or by external sources.

Molecular emission lines of H$_2$CO with the same $J_\mathrm{u}$ quantum number can be used as a probe of the kinetic temperature for cool ($T\lesssim50$~K) gas at densities  $\gtrsim10^5$~cm$^{-3}$ \citep{mangum93}. Through an unbiased survey of H$_2$CO toward embedded sources in the Corona Australis (CrA) star-forming region it was possible to characterize the temperature and external irradiation of such protostars, identifying the Herbig~Be star R~CrA as the dominant source of irradiation in CrA, heating nearby envelopes to temperatures $\sim40$~K, but also influencing the temperature on scales of a few 10\,000~au \citep{lindberg15}. The same molecular transitions were used by \citet{lindberg12} to produce an interferometric map of the physical characteristics in the R~CrA cloud (which hosts a handful of embedded low-mass protostars). The rotational temperatures of H$_2$CO were found to exceed 40~K across the whole cloud ($\sim6000$~au), consistent with \textit{Herschel}/PACS observations of the extended dust temperature \citep{lindberg14_herschel}. In particular, strong H$_2$CO and CH$_3$OH emission with high temperatures and column densities was detected toward two long ($\sim5000$~au) ridges north and south of the low-mass young stellar objects (YSOs), whereas only faint H$_2$CO emission was detected toward the YSOs themselves. On the other hand, most of the fainter \ctht\ emission was detected from the region between the two ridges. The S/N level of the observations was too low to establish \ctht\ rotational temperatures, but in single-dish APEX observations the rotational temperature of \ctht\ was found to be consistently low ($\sim10$~K) toward the embedded protostars in CrA \citep{lindberg15}.

The Ophiuchus star-forming region, at a distance of only 125~pc \citep{degeus89,wilking08}, is an excellent laboratory to test the knowledge gained from the CrA survey on a region with a greater number of potential heating sources, but also a greater separation between the low-mass sources, allowing for a distinction between effects from internal and external heating. The population of deeply embedded protostars was surveyed by \citet{jorgensen08} and \citet{enoch09} using \textit{Spitzer} and JCMT/SCUBA continuum observations, and in combination, the two surveys identified 38 Class~0 and Class~I protostars.

\begin{figure*}
	\epsscale{1.1}
	\plotone{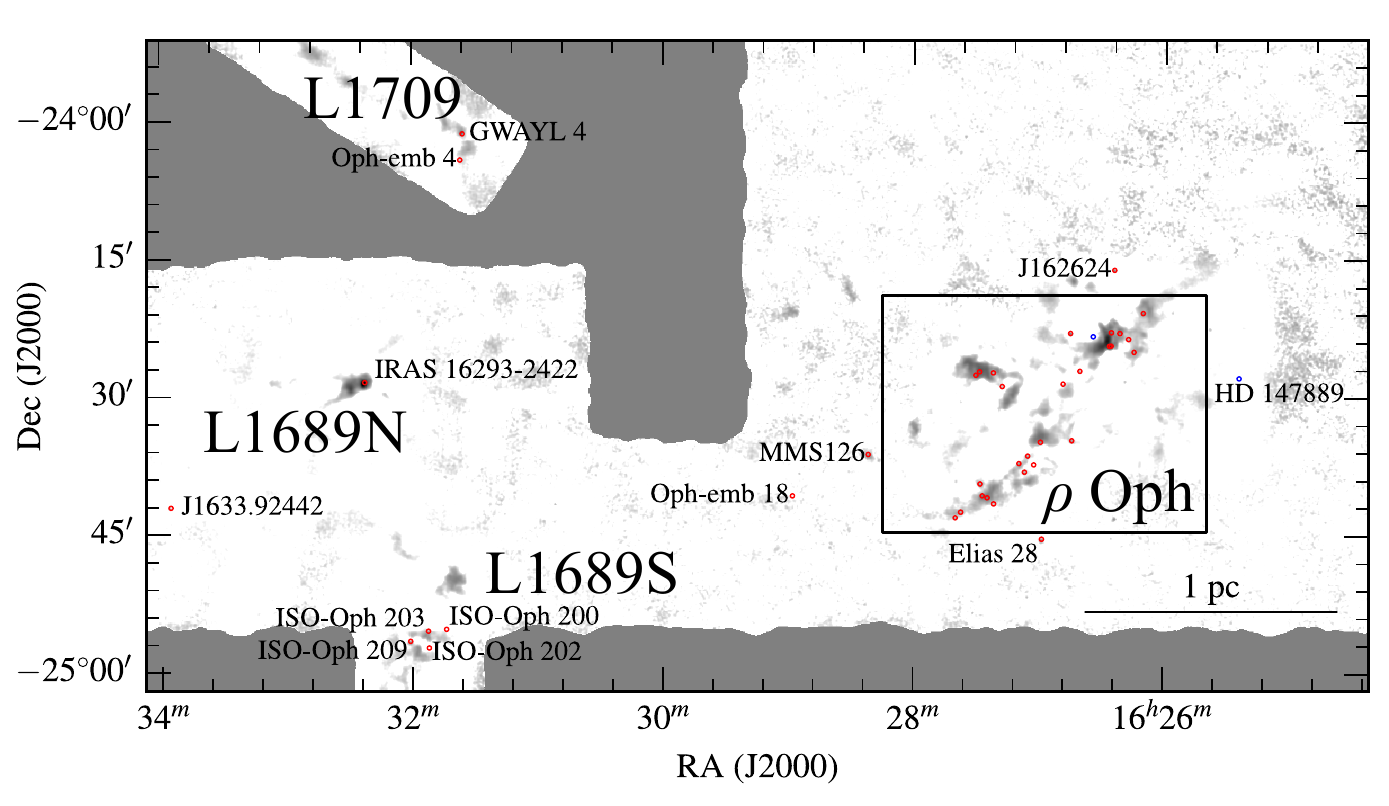}
	\caption{SCUBA 850\moo\ image of the Ophiuchus star-forming region \citep{johnstone04}. The uniformly gray areas were not covered by the SCUBA map. The sources targeted in this study are shown with red rings. In addition, the two luminous stars S~1 and HD~147889 are shown with blue rings. The area in the rectangle ($\rho$~Oph) is shown at a larger scale in \autoref{fig:zoomin}.}
	\label{fig:overview}
\end{figure*}  

\begin{figure}
	\epsscale{1.2}
	\plotone{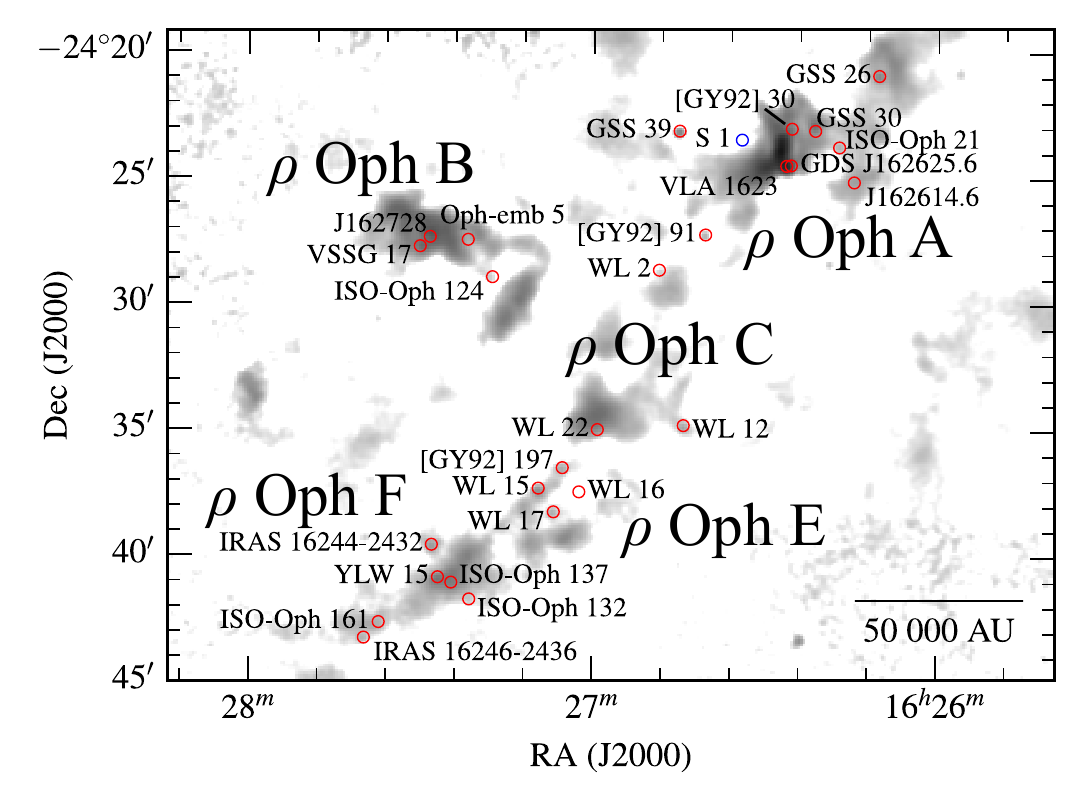}
	\caption{Zoom-in of the $\rho$~Oph cloud in \autoref{fig:overview}. The sources targeted in this study are shown with red rings. The luminous source S~1 is shown with a blue ring. The size of the rings represent the APEX beam size of 29\arcsec.}
	\label{fig:zoomin}
\end{figure}

\citet{kamegai03} found large-scale elevated [\ion{C}{1}] excitation temperatures in the $\rho$~Oph~A cloud using the Mount Fuji submillimeter telescope, and identified the B2 star HD~147889 as the heating source. \citet{liseau15}, however, concluded that HD~147889 and the B4 star S~1 (also known as GSS~35) both influence the temperature of the cloud, but that S~1, despite being less luminous, is the more dominant heating source owing to its proximity to the $\rho$~Oph~A cloud. Here, we investigate the influence these luminous stars have on the physical properties of the molecular gas in nearby embedded protostars through 218~GHz spectral line observations.

\section{Observations}

\capstartfalse
\begin{deluxetable*}{lcccrrrl}

	\tablecaption{Coordinates and other properties of the observed sources}

	\tablehead{\colhead{\textbf{Cloud}} & \colhead{R.A.} & \colhead{Dec.} & \colhead{Our rms} & \colhead{$T_\mathrm{bol}$} & \colhead{$L_\mathrm{bol}$} &  \colhead{$\alpha_\mathrm{IR}$} & \colhead{Other common} \\ 
		\colhead{Source} & \colhead{(J2000.0)} & \colhead{(J2000.0)} & \colhead{[mK~(km~s$^{-1}$)$^{-1}$]}  & \colhead{[K]} & \colhead{[$L_\odot$]} & \colhead{} & \colhead{identifiers} }

	%% All data must appear between the \startdata and \enddata commands
	\startdata
	\textbf{$\rho$ Oph A} \\
	GDS J162625.6 & 16:26:25.62 & $-24$:24:28.9 & 15.0 & $<\phantom{0}72$ & $>0.039\phantom{0}$ & $1.65$ \\
	GSS 26 & 16:26:10.33 & $-24$:20:54.8 & 26.6 & $920$ & $0.3\phantom{000}$ & $-0.46$ \\
	GSS 30 & 16:26:21.42 & $-24$:23:06.4 & 13.9 & $150$ & $8.7\phantom{000}$ & $1.46$ & Elias 21 \\
	GSS 39 & 16:26:45.03 & $-24$:23:07.7 & 24.6 & $970$ & $0.23\phantom{00}$ & $-0.64$ & Elias 27 \\
	{[}GY92{]} 30 & 16:26:25.49 & $-24$:23:01.6 & 12.0 & $135$ & $0.10\phantom{00}$ & $0.87$ \\
	{[}GY92{]} 91 & 16:26:40.47 & $-24$:27:14.5 & 16.9 & $<370$ & $>0.065\phantom{0}$ & $0.45$ & CRBR42 \\
	ISO-Oph 21 & 16:26:17.23 & $-24$:23:45.4 & 13.7 & $490$ & $0.083\phantom{0}$ & $0.69$ & CRBR12 \\
	J162614.6 & 16:26:14.62 & $-24$:25:08.4 & 14.7 & ... & ...\phantom{0000} & ... \\
	VLA 1623 & 16:26:26.42 & $-24$:24:30.0 & 16.2 & $57$ & $0.41\phantom{00}$ & $1.65$ \\
	\textbf{$\rho$ Oph B} \\
	ISO-Oph 124 & 16:27:17.57 & $-24$:28:56.3 & 15.1 & $68$ & $2.6\phantom{000}$ & $0.25$ \\
	J162728 & 16:27:28.45 & $-24$:27:21.0 & 13.3 & $310$ & $0.48\phantom{00}$ & $-0.03$ & IRS45, [GY92]~273 \\
	Oph-emb 5 & 16:27:21.83 & $-24$:27:27.6 & 13.1 & $180$ & $0.019\phantom{0}$ & $-0.05$ \\
	VSSG 17 & 16:27:30.18 & $-24$:27:43.4 & 15.2 & $530$ & $0.93\phantom{00}$ & $-0.12$ & Elias 33, IRS47, [GY92]~279 \\
	\textbf{$\rho$ Oph C} \\
	WL 2 & 16:26:48.49 & $-24$:28:38.9 & 22.6 & $430$ & $0.12\phantom{00}$ & $0.02$ & [GY92]~128 \\
	WL 12 & 16:26:44.19 & $-24$:34:48.3 & 30.5 & $290$ & $1.1\phantom{000}$ & $2.49$ & ISO-Oph 65 \\
	WL 22 & 16:26:59.17 & $-24$:34:58.8 & 14.2 & $110$ & $1.5\phantom{000}$ & $1.99$ & ISO-Oph 90 \\
	\textbf{$\rho$ Oph E} \\
	{[}GY92{]} 197 & 16:27:05.25 & $-24$:36:29.8 & 20.5 & $110$ & $0.15\phantom{00}$ & $1.27$ & LFAM26 \\
	WL 15 & 16:27:09.43 & $-24$:37:18.8 & 27.3 & $260$ & $18\phantom{.0000}$ & $1.69$ & Elias 29, [GY92]~214 \\
	WL 16 & 16:27:02.34 & $-24$:37:27.2 & 23.9 & $320$ & $4.8\phantom{000}$ & $1.53$ & [GY92]~182 \\
	WL 17 & 16:27:06.78 & $-24$:38:15.0 & 26.6 & $310$ & $0.60\phantom{00}$ & $0.61$ & [GY92]~205 \\
	\textbf{$\rho$ Oph F} \\
	IRAS 16244-2432 & 16:27:28.03 & $-24$:39:33.5 & 26.3 & $110$ & $15\phantom{.0000}$ & $2.29$ & IRS44, [GY92]~269 \\
	IRAS 16246-2436 & 16:27:39.83 & $-24$:43:15.1 & 13.6 & $570$ & $0.71\phantom{00}$ & $-0.15$ & IRS51,[GY92]~315 \\
	ISO-Oph 132 & 16:27:21.47 & $-24$:41:43.1 & 25.0 & $600$ & $1.2\phantom{000}$ & $-0.03$ & [GY92]~252 \\
	ISO-Oph 137 & 16:27:24.61 & $-24$:41:03.4 & 25.9 & $191$ & $0.33\phantom{00}$ & $1.01$ & CRBR85 \\
	ISO-Oph 161 & 16:27:37.25 & $-24$:42:38.0 & 25.9 & $450$ & $0.13\phantom{00}$ & $0.13$ & [GY92]~301 \\
	YLW 15 & 16:27:26.94 & $-24$:40:50.8 & 24.3 & $160$ & $3.8\phantom{000}$ & $1.17$ & IRS 43, [GY92]~265 \\
	\textbf{L1689N} \\
	IRAS 16293-2422 & 16:32:22.56 & $-24$:28:31.8 & 12.5 & $47$ & $16\phantom{.0000}$ & $5.03$ \\
	\textbf{L1689S} \\
	ISO-Oph 200 & 16:31:43.75 & $-24$:55:24.6 & 24.9 & $500$ & $0.28\phantom{00}$ & $0.23$ \\
	ISO-Oph 202 & 16:31:52.06 & $-24$:57:26.0 & 26.2 & $<120$ & $>0.0093$ & $0.82$ \\
	ISO-Oph 203 & 16:31:52.45 & $-24$:55:36.2 & 10.9 & $240$ & $0.13\phantom{00}$ & $1.07$ \\
	ISO-Oph 209 & 16:32:01.00 & $-24$:56:42.0 & 26.4 & $130$ & $4.0\phantom{000}$ & $1.39$ & L1689S1, IRS67 \\
	\textbf{L1709} \\
	GWAYL 4 & 16:31:35.65 & $-24$:01:29.3 & 24.1 & $300$ & $1.4\phantom{000}$ & $0.14$ & IRAS 16285-2355 \\
	Oph-emb 4 & 16:31:36.80 & $-24$:04:20.1 & 27.8 & $77$ & $0.18\phantom{00}$ & $-0.27$ \\
	\textbf{Solitary sources} \\
	Elias 28 & 16:26:58.44 & $-24$:45:31.9 & 24.1 & $860$ & $1.2\phantom{000}$ & $-0.86$ & SR 24N \\
	J162624 & 16:26:24.07 & $-24$:16:13.5 & 25.0 & $980$ & $1.9\phantom{000}$ & $-0.71$ & Elias 24 \\
	J1633.92442 & 16:33:55.61 & $-24$:42:05.0 & 23.4 & $1500$ & $0.17\phantom{00}$ & $-1.22$ \\
	MMS126 & 16:28:21.61 & $-24$:36:23.4 & 14.9 & $41$ & $0.29\phantom{00}$ & $1.23$ & IRAS 16253-2429 \\
	Oph-emb 18 & 16:28:57.85 & $-24$:40:54.9 & 19.7 & $300$ & $0.03\phantom{00}$ & $0.67$

	\enddata
	
	%% Include any \tablenotetext{key}{text}, \tablerefs{ref list},
	%% or \tablecomments{text} between the \enddata and 
	%% \end{deluxetable} commands
	
	%% No \tablecomments indicated
	
	%% No \tablerefs indicated
	\tablecomments{Spectral properties from \citet{evans09}, except for the values of {[}GY92{]} 30, VLA~1623, and ISO-Oph~137, which are from \citet{enoch09}.}
	
	\label{tab:sourcelist}
\end{deluxetable*}
\capstarttrue

Our observations were executed with the APEX (Atacama Pathfinder Experiment) 12 m telescope \citep{gusten06} in position-switching mode in August and October 2014 and May 2015. The SHeFI APEX-1 receiver \citep{vassilev08} was used to cover the frequency range 216--220~GHz toward a complete sample of the 38 deeply embedded protostellar sources in the Ophiuchus star-forming region. The sample consists of all embedded protostellar sources in the region detected by \citet{jorgensen08} and \citet{enoch09}, see \autoref{tab:sourcelist} and Figures~\ref{fig:overview}--\ref{fig:zoomin}. The spectral setup was chosen to cover three H$_2$CO spectral lines at 218~GHz, and also several spectral lines of \ctht, an abundant tracer of unsaturated hydrocarbon molecules.

We note that five of the sources in the sample (GSS~26, GSS~39, Elias~28, J162624, and J1633.92442) should be considered Class~II sources by both the $T_\mathrm{bol}>650$~K and the $\alpha_\mathrm{IR}<-0.3$ criteria \citep{evans09}, but are still included in the sample of embedded protostars of \citet{jorgensen08} due to their proximity ($<15\arcsec$) to a SCUBA core. We detect no line emission toward one of these sources (J1633.92442) and only C$^{18}$O line emission toward two (GSS~39 and Elias~28), and these sources are thus likely not deeply embedded. GSS~26 and J162624, however, show H$_2$CO line emission, indicating the presence of dense molecular gas. Finally, toward the two Class~I sources WL16 and Oph-emb~4, we only detect emission from C$^{18}$O.

To remove quasi-sinusoidal baselines present in the APEX-1 spectra \citep[see][]{vassilev08}, we used our running-mean script described in Appendix~A of \citet{lindberg15}. The spectra were thereafter reduced with the X-Spec package\footnote{X-Spec is developed at Onsala Space observatory, and can be obtained from \url{http://www.chalmers.se/en/centres/oso/radio-astronomy/Pages/ software.aspx}}, which was used to average scans, perform Gaussian fits, and calculate line intensities. Throughout the paper we use the $T_\mathrm{mb}$ temperature scale, assuming a main beam efficiency $\eta_\mathrm{mb}=0.75$.

\section{Spectral line profiles}

The spectral lines which were used in the analysis are listed in \autoref{tab:spectrallines}. These lines were not detected toward all sources, and in addition to these lines, several other lines (of e.g. C$_2$D, DCN, C$^{18}$O, C$^{33}$S, and HNCO) were detected in some of the sources, but will not be discussed in this paper. All observed line parameters are listed in \autoref{tab:obslines} in \autoref{app:obslines}.

\capstartfalse
\begin{deluxetable}{lllll}
	
	%% Keep a portrait orientation
	
	%% Over-ride the default font size
	%% Use Default (12pt)
	
	%% Use \tablewidth{?pt} to over-ride the default table width.
	%% If you are unhappy with the default look at the end of the
	%% *.log file to see what the default was set at before adjusting
	%% this value.
	
	%% This is the title of the table.
	\tablecaption{Spectral lines used in the analysis}
	
	%% This command over-rides LaTeX's natural table count
	%% and replaces it with this number.  LaTeX will increment 
	%% all other tables after this table based on this number
	%\tablenum{2}
	
	%% The \tablehead gives provides the column headers.  It
	%% is currently set up so that the column labels are on the
	%% top line and the units surrounded by ()s are in the 
	%% bottom line.  You may add more header information by writing
	%% another line between these lines. For each column that requries
	%% extra information be sure to include a \colhead{text} command
	%% and remember to end any extra lines with \\ and include the 
	%% correct number of &s.
	\tablehead{\colhead{Molecule} & \colhead{Quantum } & \colhead{Frequency} & \colhead{Type} & $E_\mathrm{u}/k$ \\ 
		\colhead{} & \colhead{numbers} & \colhead{[GHz]} & & [K] } 
	
	%% All data must appear between the \startdata and \enddata commands
	\startdata
DCO$^+$ & $3\rightarrow2$ & 216.11258 & ... & 20.7 \\
\ctht\ & $3_{03}\rightarrow2_{02}$ & 216.27876  & ortho & 19.5 \\
\ctht\ & $6_{06}\rightarrow5_{15}$ & 217.82215  & para  & 38.6 \\
\ctht\ & $6_{16}\rightarrow5_{05}$ & 217.82215  & ortho & 38.6 \\
\ctht\ & $5_{14}\rightarrow4_{23}$ & 217.94005  & ortho & 35.4 \\
\ctht\ & $5_{24}\rightarrow4_{13}$ & 218.16046  & para  & 35.4 \\
H$_2$CO                    & $3_{03}\rightarrow2_{02}$ & 218.22219  & para  & 21.0 \\ 
CH$_3$OH                   & $4_{2}\rightarrow3_{1}$   & 218.44006  & E     & 45.5 \\
H$_2$CO                    & $3_{22}\rightarrow2_{21}$ & 218.47563  & para  & 68.1 \\
\ctht\ & $7_{16}\rightarrow7_{07}$ & 218.73273  & ortho & 61.2 \\
\ctht\ & $7_{26}\rightarrow7_{17}$ & 218.73273  & para  & 61.2 \\
H$_2$CO                    & $3_{12}\rightarrow2_{11}$ & 218.76007  & para  & 68.1 \\
SO & $6_5\rightarrow5_4$ & 219.94944 & ... & 35.0

\enddata

%% Include any \tablenotetext{key}{text}, \tablerefs{ref list},
%% or \tablecomments{text} between the \enddata and 
%% \end{deluxetable} commands

%% No \tablecomments indicated

%% No \tablerefs indicated
\tablecomments{Rest frequencies from the CDMS \citep{cdms} and JPL \citep{jpl} molecular spectroscopy databases.}

\label{tab:spectrallines}
\end{deluxetable}
\capstarttrue

\begin{figure*}
	\epsscale{1.2}
	\plotone{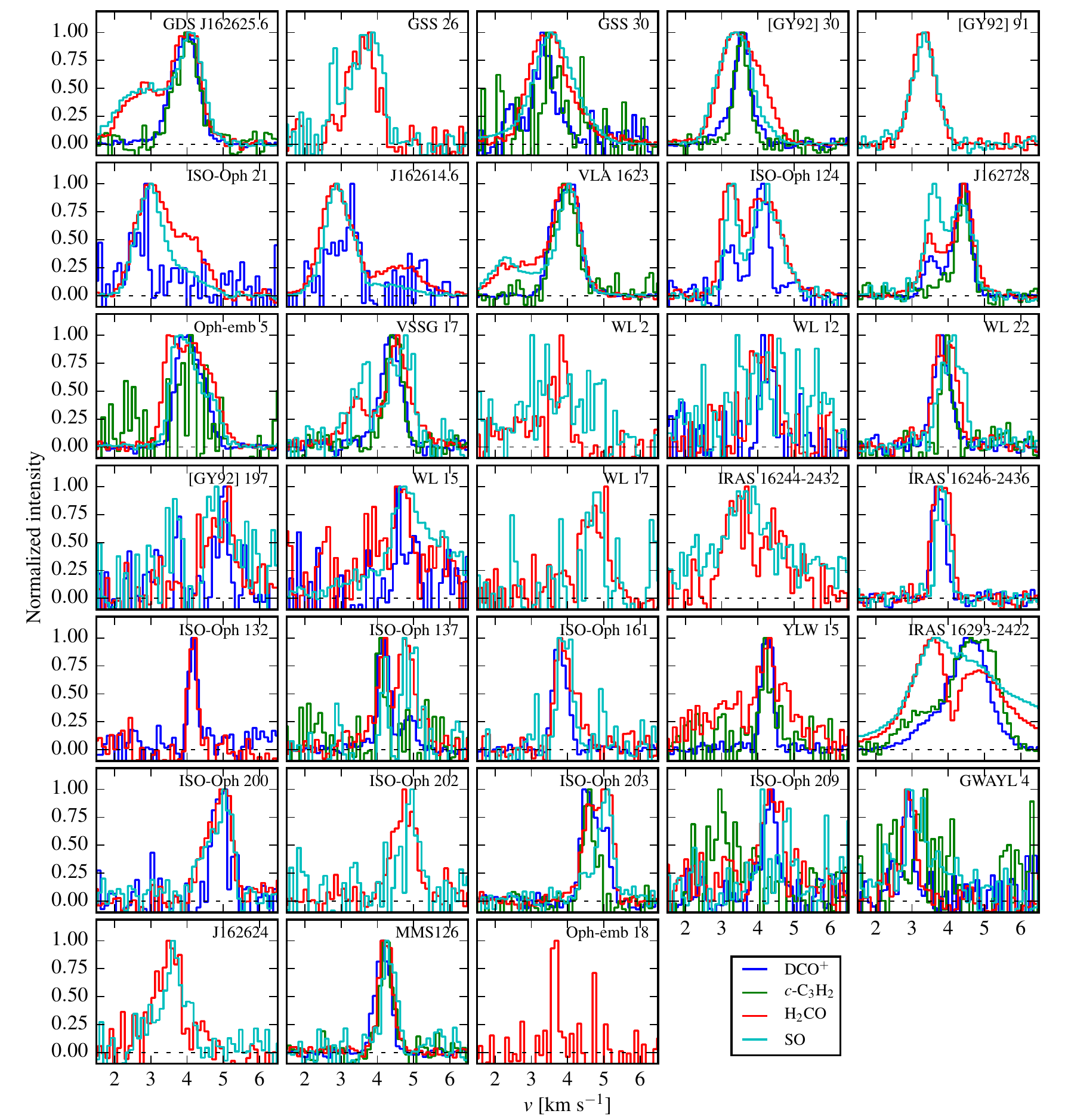}
	\caption{Spectra of the DCO$^+$ (216.113~GHz), \ctht\ (217.822~GHz), H$_2$CO (218.222~GHz), and SO (219.949~GHz) lines for all sources in the sample for which at least one of these lines is detected at a $3\sigma$ level or higher (Elias~28, GSS~39, J1633.9-2442, Oph-emb~4, and WL~16 are not included since no such lines are detected), normalized to their individual peaks to facilitate comparison of the line profiles.}
	\label{allspectra}
\end{figure*}  

To enable comparison of the line profiles, \autoref{allspectra} shows the normalized spectra for the strongest DCO$^+$ (216.113~GHz), \ctht\ (217.822~GHz), H$_2$CO (218.222~GHz), and SO (219.949~GHz) lines in each source where they have been detected. We note that for some sources, the line profiles of different molecular species are dramatically discrepant. In several cases, a double peak is seen in some spectral lines. The CH$_3$OH line (not plotted) is generally well-correlated with the H$_2$CO line shape, while C$_2$D (not plotted) correlates well with \ctht, although the C$_2$D line doublet makes analysis of the line profile non-trivial. Of the plotted lines, we find that the line profiles often separate the molecular species into two groups: H$_2$CO and SO on one hand, and \ctht\ and DCO$^+$ on the other. \citet{lindberg15} found that H$_2$CO and other saturated organic species as well as SO and other sulfur-bearing species had significantly higher rotational temperatures than \ctht\ and other unsaturated hydrocarbons in the APEX observations of the embedded protostar R~CrA IRS7B. Like in CrA, the H$_2$CO and SO line profiles are generally wider than the \ctht\ line profiles. This is also similar to the six molecular cores studied by \citet{buckle06}, where CH$_3$OH and SO were found to be spatially correlated, but separated from the unsaturated hydrocarbon HC$_3$N.

For completeness, the corresponding spectra for the Corona Australis survey are shown in \autoref{fig:allspectra_cra} of \autoref{sec:cra}.

\subsection{Secondary components}
\label{sec:secondary}

In at least nine sources (four in $\rho$~Oph~A, three in $\rho$~Oph~B, one in L1689S, and IRAS~16293\nobr2422), a secondary component to some of the emission lines is detected. This was also seen in two sources in the CrA survey: SMM2 and CrA-24. In all eleven cases, the additional component is characterized by strong H$_2$CO and CH$_3$OH emission, and often also SO, but faint or no emission from other species. In CrA, the component likely comes from an overlapping molecular outflow (Miettinen et~al., in~prep.). We note that the velocity of the secondary component is blueward of the main component in all $\rho$~Oph~A and $\rho$~Oph~B sources with the exception of ISO-Oph~21 and J162614.6.

To account for the possibly distinct nature of these secondary velocity components, they are treated separately from the main component in the rotational temperature analysis, and are plotted as red data points in all plots (although they are typically not the redward components). We identify which component to be considered ``secondary'' by investigating the molecular emission prevalent at those velocities -- typically, the secondary component has little or no DCO$^+$ and \ctht\ emission. The secondary component is often also wider than the main component, and usually appears at a velocity different from the typical velocity of other sources in the same cloud (thus possibly representing a molecular outflow). For the rotational diagram analysis (see below), emission from the two components were separated by performing Gaussian fits. We find that the secondary component in most cases has a higher H$_2$CO rotational temperature than the main component, in agreement with the CrA sources.

\begin{figure*}
	\epsscale{1.1}
	\plottwo{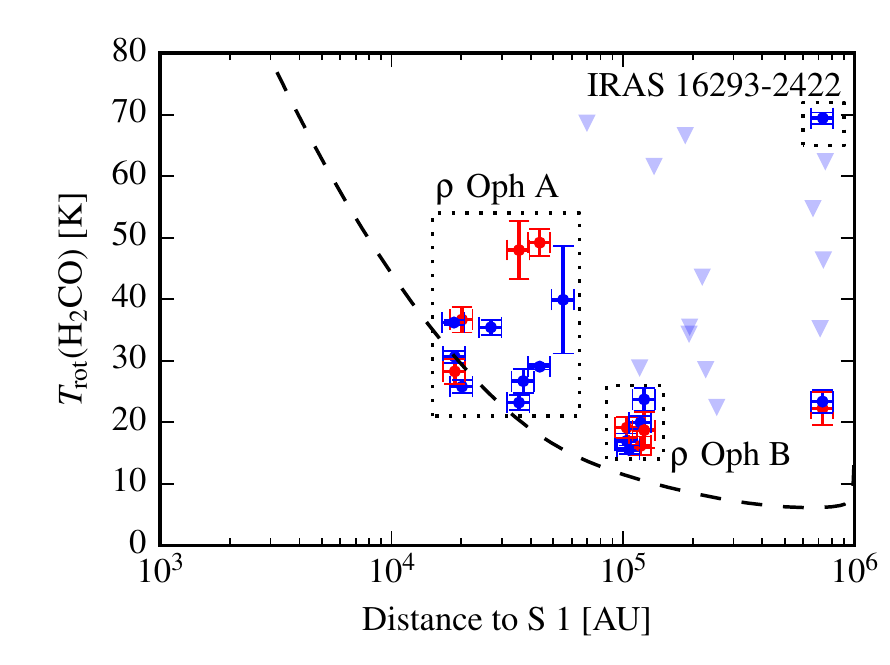}{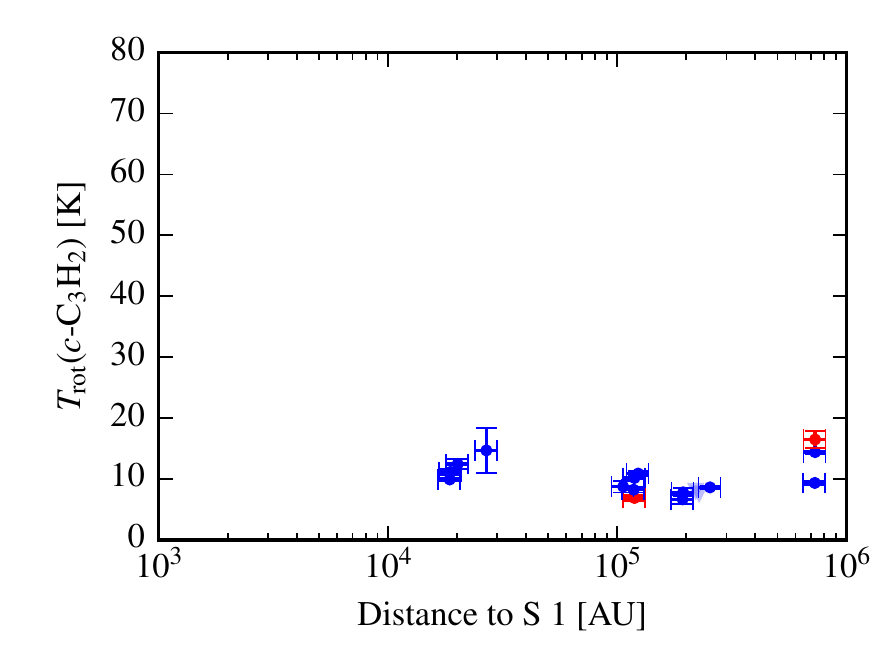}
	\caption{\textit{Left:} H$_2$CO rotational temperature as function of the projected distance to the Herbig~Be star S~1. Red data points represent the secondary component dominated by H$_2$CO, CH$_3$OH, and SO in sources with two line components, (see \autoref{sec:secondary}). The secondary component of IRAS~16293\nobr2422 is off the scale, at 124~K. The dotted rectangles show sources belonging to the clouds $\rho$~Oph~A and $\rho$~Oph~B, and the \textit{hot corino} IRAS~16293\nobr2422. The black dashed curve shows a \textit{Transphere} \citep{transphere} model of the dust temperature resulting from the heating from S~1. The heating from the more luminous, although more distant, HD~147889 is not included since its contribution to the heating of most sources is minor in comparison with S~1 (see \autoref{fig:contributions}). \textit{Right:} \ctht\ rotational temperature as function of the projected distance to the Herbig~Be star S~1. In both figures, light-blue triangles are upper limits.}
	\label{h2co_temp}
\end{figure*}

\section{Rotational temperatures}
\label{sec:rottemp}

\capstartfalse
\begin{deluxetable*}{lllllllll}

	\tablecaption{Measured rotational temperatures, column densities, median line widths, and LSR velocities}

	\tablehead{
	\colhead{} &\multicolumn4c{H$_2$CO} & \multicolumn4c{\ctht}\\
	\colhead{\textbf{Cloud}} & \colhead{$T_\mathrm{rot}$} & \colhead{$N$} & \colhead{$\Delta v$} & \colhead{$v_\mathrm{LSR}$} & \colhead{$T_\mathrm{rot}$} & \colhead{$N_\mathrm{rot}$} & \colhead{$\Delta v$} & \colhead{$v_\mathrm{LSR}$}\\ 
	\colhead{Source} & \colhead{[K]} & \colhead{[$10^{12}$ cm$^{-2}$]} & \colhead{[km~s$^{-1}$]} & \colhead{[km~s$^{-1}$]} & \colhead{[K]} & \colhead{[$10^{12}$ cm$^{-2}$]} & \colhead{[km~s$^{-1}$]} & \colhead{[km~s$^{-1}$]}} 
	
	%% All data must appear between the \startdata and \enddata commands
	\startdata
	\textbf{$\rho$ Oph A} \\
	GDS J162625.6 (main) & $\phantom{00}\phantom{0}25.8\pm1.0$ & $18\phantom{.00}\pm3\phantom{.00}$ & 0.8 & 3.8  & $\phantom{00}12.4\pm0.8$ & $\phantom{0}2.5\phantom{0}\pm0.4\phantom{0}$ & 0.8 & 3.8  \\
	GDS J162625.6 (sec.) & $\phantom{00}\phantom{0}36.7\pm2.1$ & $11\phantom{.00}\pm1\phantom{.00}$ & 1.4 & 2.6  & \phantom{000}...  & \phantom{0}...  & ... & ... \\
	GSS 26 & $\phantom{00}\phantom{0}39.9\pm8.7$ & $\phantom{0}4.3\phantom{0}\pm1.4\phantom{0}$ & 1.1 & 3.4  & \phantom{000}...  & \phantom{0}...  & ... & ... \\
	GSS 30 & $\phantom{00}\phantom{0}35.4\pm1.2$ & $19\phantom{.00}\pm2\phantom{.00}$ & 1.3 & 3.3  & $\phantom{00}14.7\pm3.7$ & $\phantom{0}0.46\pm0.26$ & 0.9 & 3.5  \\
	GSS 39 & \phantom{0000}... & \phantom{0}...  & ... & ... & \phantom{000}...  & \phantom{0}...  & ... & ... \\
	{[}GY92{]} 30 & $\phantom{00}\phantom{0}36.2\pm0.4$ & $79\phantom{.00}\pm8\phantom{.00}$ & 1.2 & 3.4  & $\phantom{00}\phantom{0}9.9\pm0.3$ & $\phantom{0}4.0\phantom{0}\pm0.3\phantom{0}$ & 0.4 & 3.4  \\
	{[}GY92{]} 91 & $\phantom{00}\phantom{0}26.7\pm2.0$ & $12\phantom{.00}\pm2\phantom{.00}$ & 0.6 & 3.2  & \phantom{000}...  & \phantom{0}...  & ... & ... \\
	ISO-Oph 21 (main) & $\phantom{00}\phantom{0}23.2\pm1.2$ & $15\phantom{.00}\pm3\phantom{.00}$ & 0.9 & 2.8  & \phantom{000}...  & \phantom{0}...  & ... & ... \\
	ISO-Oph 21 (sec.) & $\phantom{00}\phantom{0}48.0\pm4.7$ & $\phantom{0}4.4\phantom{0}\pm0.6\phantom{0}$ & 1.2 & 3.9  & \phantom{000}...  & \phantom{0}...  & ... & ... \\
	J162614.6 (main) & $\phantom{00}\phantom{0}29.0\pm0.5$ & $47\phantom{.00}\pm6\phantom{.00}$ & 0.9 & 2.7  & \phantom{000}...  & \phantom{0}...  & ... & ... \\
	J162614.6 (sec.) & $\phantom{00}\phantom{0}49.2\pm2.2$ & $11\phantom{.00}\pm1\phantom{.00}$ & 1.5 & 4.3  & \phantom{000}...  & \phantom{0}...  & ... & ... \\
	VLA 1623 (main) & $\phantom{00}\phantom{0}30.6\pm1.0$ & $23\phantom{.00}\pm3\phantom{.00}$ & 0.8 & 3.8  & $\phantom{00}11.2\pm0.5$ & $\phantom{0}4.2\phantom{0}\pm0.5\phantom{0}$ & 0.8 & 3.8  \\
	VLA 1623 (sec.) & $\phantom{00}\phantom{0}28.3\pm2.1$ & $12\phantom{.00}\pm2\phantom{.00}$ & 1.4 & 2.5  & \phantom{000}...  & \phantom{0}...  & ... & ... \\
	\textbf{$\rho$ Oph B} \\
	ISO-Oph 124 (main) & $\phantom{00}\phantom{0}17.0\pm1.2$ & $23\phantom{.00}\pm5\phantom{.00}$ & 1.1 & 4.3  & \phantom{000}...  & \phantom{0}...  & ... & ... \\
	ISO-Oph 124 (sec.) & $\phantom{00}\phantom{0}19.1\pm1.7$ & $\phantom{0}9.2\phantom{0}\pm2.4\phantom{0}$ & 0.5 & 3.1  & \phantom{000}...  & \phantom{0}...  & ... & ... \\
	J162728 (main) & $\phantom{00}\phantom{0}20.0\pm1.1$ & $21\phantom{.00}\pm4\phantom{.00}$ & 0.6 & 4.2  & $\phantom{00}10.2\pm0.2$ & $\phantom{0}4.1\phantom{0}\pm0.3\phantom{0}$ & 0.4 & 4.3  \\
	J162728 (sec.) & $\phantom{00}\phantom{0}16.2\pm1.5$ & $13\phantom{.00}\pm4\phantom{.00}$ & 0.5 & 3.4  & $\phantom{00}\phantom{0}6.9\pm0.5$ & $\phantom{0}1.9\phantom{0}\pm0.5\phantom{0}$ & 0.5 & 3.7  \\
	Oph-emb 5 & $\phantom{00}\phantom{0}15.5\pm0.7$ & $55\phantom{.00}\pm\phantom{}12\phantom{.00}$ & 0.8 & 3.7  & $\phantom{00}\phantom{0}8.7\pm1.0$ & $\phantom{0}1.0\phantom{0}\pm0.4\phantom{0}$ & 0.5 & 3.9  \\
	VSSG 17 (main) & $\phantom{00}\phantom{0}23.7\pm1.8$ & $12\phantom{.00}\pm2\phantom{.00}$ & 0.8 & 4.4  & $\phantom{00}10.9\pm0.4$ & $\phantom{0}3.2\phantom{0}\pm0.4\phantom{0}$ & 0.5 & 4.3  \\
	VSSG 17 (sec.) & $\phantom{00}\phantom{0}18.7\pm2.9$ & $\phantom{0}6.2\phantom{0}\pm2.4\phantom{0}$ & 0.5 & 3.3  & \phantom{000}...  & \phantom{0}...  & ... & ... \\
	\textbf{$\rho$ Oph C} \\
	WL 2 & $<120\phantom{.0}$ & ...  & 0.5 & 3.7  & \phantom{000}...  & \phantom{0}...  & ... & ... \\
	WL 12 & $<215\phantom{.0}$ & ...  & 0.9 & 4.0  & \phantom{000}...  & \phantom{0}...  & ... & ... \\
	WL 22 & $<\phantom{0}28.8$ & ...  & 0.8 & 3.8  & $\phantom{00}\phantom{0}8.2\pm0.5$ & $\phantom{0}2.4\phantom{0}\pm0.5\phantom{0}$ & 0.4 & 3.8  \\
	\textbf{$\rho$ Oph E} \\
	{[}GY92{]} 197 & $<\phantom{0}61.6$ & ...  & 0.9 & 4.8  & \phantom{000}...  & \phantom{0}...  & ... & ... \\
	WL 15 & $<109\phantom{.0}$ & ...  & 2.2 & 4.5  & \phantom{000}...  & \phantom{0}...  & ... & ... \\
	WL 16 & \phantom{0000}... & \phantom{0}...  & ... & ... & \phantom{000}...  & \phantom{0}...  & ... & ... \\
	WL 17 & $<292\phantom{.0}$ & ...  & 0.9 & 4.6  & \phantom{000}...  & \phantom{0}...  & ... & ... \\
	\textbf{$\rho$ Oph F} \\
	IRAS 16244-2432 & $<\phantom{0}66.6$ & ...  & 1.9 & 3.7  & \phantom{000}...  & \phantom{0}...  & ... & ... \\
	IRAS 16246-2436 & $<\phantom{0}28.5$ & ...  & 0.6 & 3.7  & $<\phantom{0}7.5$ & ...  & 0.3 & 3.7  \\
	ISO-Oph 132 & $<138\phantom{.0}$ & ...  & 0.3 & 4.0  & \phantom{000}...  & \phantom{0}...  & ... & ... \\
	ISO-Oph 137 & $<\phantom{0}34.3$ & ...  & 1.3 & 4.4  & $\phantom{00}\phantom{0}6.7\pm0.8$ & $\phantom{0}2.5\phantom{0}\pm1.0\phantom{0}$ & 0.2 & 4.0  \\
	ISO-Oph 161 & $<\phantom{0}43.6$ & ...  & 0.6 & 3.8  & $<\phantom{0}7.9$ & ...  & 0.4 & 3.7  \\
	YLW 15 & $<\phantom{0}35.4$ & ...  & 2.4 & 3.9  & $\phantom{00}\phantom{0}7.8\pm0.7$ & $\phantom{0}2.6\phantom{0}\pm0.8\phantom{0}$ & 0.4 & 4.1  \\
	\textbf{L1689N} \\
	IRAS 16293-2422 (main) & $\phantom{00}\phantom{0}69.4\pm0.9$ & $47\phantom{.00}\pm3\phantom{.00}$ & 2.7 & 4.8  & $\phantom{00}14.4\pm0.3$ & $\phantom{0}6.5\phantom{0}\pm0.3\phantom{0}$ & 1.3 & 4.6  \\
	IRAS 16293-2422 (sec.) & $\phantom{00}124\phantom{.0}\pm3\phantom{.0}$ & $32\phantom{.00}\pm1\phantom{.00}$ & 1.7 & 3.2  & $\phantom{00}16.5\pm1.5$ & $\phantom{0}2.4\phantom{0}\pm0.5\phantom{0}$ & 1.7 & 3.1  \\
	\textbf{L1689S} \\
	ISO-Oph 200 & $<\phantom{0}35.2$ & ...  & 0.8 & 4.7  & \phantom{000}...  & \phantom{0}...  & ... & ... \\
	ISO-Oph 202 & $<\phantom{0}46.4$ & ...  & 0.8 & 4.6  & \phantom{000}...  & \phantom{0}...  & ... & ... \\
	ISO-Oph 203 (main) & $\phantom{00}\phantom{0}23.4\pm1.9$ & $\phantom{0}5.9\phantom{0}\pm1.2\phantom{0}$ & 0.5 & 4.5  & $\phantom{00}\phantom{0}9.3\pm0.4$ & $\phantom{0}1.9\phantom{0}\pm0.3\phantom{0}$ & 0.4 & 4.4  \\
	ISO-Oph 203 (sec.) & $\phantom{00}\phantom{0}22.3\pm2.7$ & $\phantom{0}4.0\phantom{0}\pm1.1\phantom{0}$ & 0.4 & 4.9  & \phantom{000}...  & \phantom{0}...  & ... & ... \\
	ISO-Oph 209 & $<\phantom{0}62.3$ & ...  & 1.3 & 4.2  & \phantom{000}...  & \phantom{0}...  & 1.3 & 2.9  \\
	\textbf{L1709} \\
	GWAYL 4 & $<\phantom{0}54.7$ & ...  & 0.5 & 2.8  & \phantom{000}...  & \phantom{0}...  & 1.3 & 3.0  \\
	Oph-emb 4 & \phantom{0000}... & \phantom{0}...  & ... & ... & \phantom{000}...  & \phantom{0}...  & ... & ... \\
	\textbf{Solitary sources} \\
	Elias 28 & \phantom{0000}... & \phantom{0}...  & ... & ... & \phantom{000}...  & \phantom{0}...  & ... & ... \\
	J162624 & $<\phantom{0}68.6$ & ...  & 0.8 & 3.3  & \phantom{000}...  & \phantom{0}...  & ... & ... \\
	J1633.92442 & \phantom{0000}... & \phantom{0}...  & ... & ... & \phantom{000}...  & \phantom{0}...  & ... & ... \\
	MMS126 & $<\phantom{0}22.4$ & ...  & 0.4 & 4.1  & $\phantom{00}\phantom{0}8.6\pm0.3$ & $\phantom{0}4.4\phantom{0}\pm0.5\phantom{0}$ & 0.4 & 4.1  \\
	Oph-emb 18 & \phantom{0000}... & \phantom{0}...  & 0.2 & 3.5  & \phantom{000}...  & \phantom{0}...  & ... & ...

	\enddata
	
	%% Include any \tablenotetext{key}{text}, \tablerefs{ref list},
	%% or \tablecomments{text} between the \enddata and 
	%% \end{deluxetable} commands
	
	%% No \tablecomments indicated
	
	%% No \tablerefs indicated
	\tablecomments{The H$_2$CO column densities were calculated by non-LTE methods assuming $n(\mathrm{H}_2) = (8.9\pm1.0)\times10^5$~cm$^{-3}$ (see \autoref{sec:rottemp}). The linewidths and LSR velocities are median values. \autoref{sec:secondary} describes the main and secondary components.
	%For the sources where main and secondary components are noted, this refers to two Gaussian components of the spectral lines (see Section~\ref{sec:secondary}).
	}
	
	\label{tab:rotdiagresults}
\end{deluxetable*}
\capstarttrue

For all sources where at least two H$_2$CO or \ctht\ spectral lines were detected, the respective rotational temperatures (assuming LTE) have been calculated. We assume that the lines are optically thin, which is valid at the observed column densities \citep{lindberg15}. For sources with only one detected line of a certain species, an upper limit of the rotational temperature was calculated by use of upper limits on the non-detected line(s). The results are presented in \autoref{tab:rotdiagresults}.

The three observed H$_2$CO lines all have the same $J_\mathrm{u} = 3$, and their rotational temperature is thus an excellent proxy for the kinetic temperature, but makes estimates of the column density more uncertain \citep{mangum93,lindberg15}. We estimate the beam-averaged H$_2$CO column densities by use of the non-LTE radiative transfer code RADEX \citep{radex} assuming an H$_2$ number density of $(8.9\pm1.0)\times10^5$~cm$^{-3}$ in all sources. This value was measured from observations of multiple H$_2$CO $J_\mathrm{u} = 3$ and $J_\mathrm{u} = 5$ lines toward the protostar R~CrA IRS7B using non-LTE methods (RADEX), and is of the same order of magnitude as found in the other protostars in CrA where this value could be measured \citep{lindberg15}. All observed H$_2$CO lines are para lines, and an ortho-to-para ratio is therefore not important for the excitation analysis. However, such a ratio is necessary to calculate the total H$_2$CO column density, and for this purpose an ortho/para ratio of 1.6 is assumed \citep[see][]{dickens99,jorgensen05b}. For the \ctht\ excitation analysis, an ortho/para ratio of 3 is assumed \citep[see][]{lucas00}. Non-LTE RADEX models of \ctht\ show that its physical temperature is well-represented by the rotational temperature if $n(\mathrm{H}_2) \gtrsim 10^5$~cm$^{-3}$, at temperatures typical for protostellar envelopes. This is further discussed in \autoref{sec:c3h2} \citep[see also][]{spezzano15}.

The H$_2$CO rotational temperatures in the Ophiuchus star-forming region are generally higher than what is expected in low-mass protostars on the scale of the APEX primary beam (29\arcsec, or 3600~au at a distance of 125~pc). Internal heating by low-mass embedded protostars would produce temperatures $\sim10$--15~K on these scales \citep{bergin07}. The H$_2$CO temperatures observed toward most sources thus require an external radiation field to be explained \citep[see e.g.][]{jorgensen06, lindberg12}.

\subsection{IRAS 16293\nobr2422}

The highest H$_2$CO rotational temperature in the survey was measured toward the Class~0 protostar IRAS~16293\nobr2422 ($69\pm1$~K in the main component and $124\pm4$~K in the secondary component). This source is the best studied example of a \textit{hot corino} \citep[see e.g.][]{cazaux03,caux11}, and as shown by interferometric observations, most of the line emission from organic species originates in the hot inner envelope ($R\lesssim100$~au), where the central object heats the molecular gas to high temperatures \citep[e.g.][]{bisschop08}. 
In contrast to the other sources in the survey (see below), which are not known to exhibit \textit{hot corino} properties, the H$_2$CO temperature measured toward IRAS~16293\nobr2422 likely traces the properties of this hot and dense inner envelope. Most other sources in the sample are also at a later stage of evolution than IRAS~16293\nobr2422 as shown by their higher bolometric temperature.

However, the \ctht\ temperatures measured toward IRAS~16293\nobr2422 are much lower, only $14.4\pm0.3$~K and $16.5\pm1.5$~K for the main and secondary components, respectively, marginally higher than the values found in the other sources in the survey, thus likely tracing emission on larger scales.

\subsection{External heating of sources in $\rho$~Oph~A and $\rho$~Oph~B}

The H$_2$CO rotational temperatures are somewhat higher toward the sources in the $\rho$~Oph~A cloud than in the $\rho$~Oph~B cloud. With two exceptions (IRAS~16293\nobr2422  and ISO\nobr Oph~203), all sources where the H$_2$CO temperatures could be measured are located in these two clouds, and it is therefore not possibe to judge whether the H$_2$CO temperatures of these clouds are representative of protostellar envelopes in the whole region.

The luminous Herbig~Be star S~1 is located only 2\arcmin~($\sim 15\,000$~au) east of $\rho$~Oph~A. It has spectral class B4 \citep{bouvier92} and a luminosity $\sim 1000$--$1600L_\odot$ \citep{bontemps01,wilking05}. About 15\arcmin~(0.5~pc) west of the cloud lies the even more luminous B2 star HD~147889 \citep{houk88}, with a luminosity $\sim4500L_\odot$ \citep{greene89}.
In their radiative transfer models, \citet{liseau15} use luminosities of S~1 and HD~147889 of $1100L_\odot$ and $4500L_\odot$, respectively.

In \autoref{h2co_temp}, the rotational temperatures of H$_2$CO and \ctht\ are plotted as a function of the projected distance to S~1. We also calculate how the luminous source S~1 influences the dust temperature, and thus indirectly the molecular gas temperature, using a 1-D \textit{Transphere} model \citep{transphere}. We assume the luminosity of S~1 to be $1600L_\odot$ and a constant cloud density $n(\mathrm{H}_2)=10^4$~cm$^{-3}$. The model (dashed line in \autoref{h2co_temp}) underpredicts the observed H$_2$CO temperatures of the sources in $\rho$~Oph~A and $\rho$~Oph~B somewhat, but the irradiation from HD~147889 and the internal irradiation from the embedded sources are not included in the model, which could account for the difference. A density different from our assumption, or a non-uniform density distribution within the cloud could also play a role in explaining the discrepancy.

\begin{figure}
	\epsscale{1.15}
	\plotone{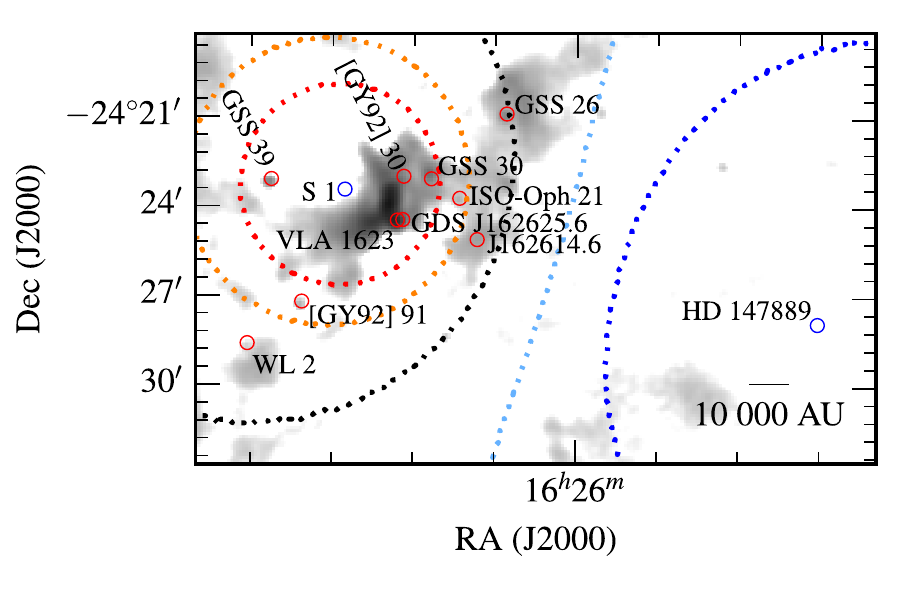}
	\caption{The black contour shows where the heating contributions of S~1 and HD~147889 are equal given our 1-D \textit{Transphere} model assuming that all sources lie in the same plane of sky. The orange and red contours show the positions where the heating from only S~1 would produce temperatures 5~K and 10~K greater than the heating from only HD~147889. The light blue and dark blue contours show the corresponding positions for HD~147889.}
	\label{fig:contributions}
\end{figure}  

To compare the contribution from the two heating sources, and to find the dominant source of heating in any given part of the cloud, \autoref{fig:contributions} shows where the heating from both sources are equal (black contour). This plot was made with two important assumptions limiting the level of interpretations to be made, namely that the cloud is uniformly dense ($n(\mathrm{H}_2)=10^4$~cm$^{-3}$) and that all sources lie in the same plane of sky.

\subsection{Temperatures of H$_2$CO and \ctht}

The H$_2$CO rotational temperature in the protostellar envelopes is strongly enhanced in the sources located close to the Herbig~Be star S~1 and the very luminous star HD~147889, whereas the \ctht\ rotational temperature is around 10--15~K in all sources (similar to the rotational temperatures of unsaturated hydrocarbon species observed in so-called warm carbon-chain chemistry sources, a type of deeply embedded protostars with high abundances of unsaturated carbon-chain molecules; see e.g.\ \citealt{sakai13}). The same trend is found in a similar study in the CrA star-forming region \citep{lindberg15}. For some reason, the H$_2$CO gas is more prone to be heated by external radiation fields, or the H$_2$CO abundance is enhanced in the irradiated gas, possibly due to photo-chemistry \citep[see e.g][]{guzman11}. We propose that the H$_2$CO emission originates in relatively outer regions of the envelope (on scales of $R\sim2000$~au; toward the edge of the APEX beam) more exposed to the external irradiation field, while the \ctht\ emission originates in inner regions of the envelope, where it is shielded against the external irradiation. This is consistent with interferometric observations of the externally irradiated protostar R~CrA IRS7B, which show heated H$_2$CO in large structures a few 1000~au from the central low-mass objects, while the \ctht\ is found closer to the protostars \citep{lindberg12}.

\begin{figure}
	\epsscale{1.1}
	\plotone{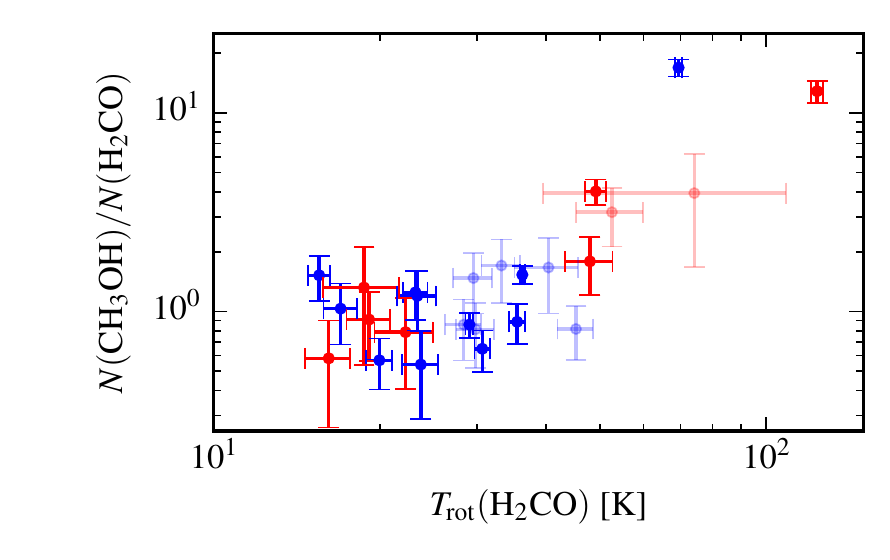}
	\caption{The CH$_3$OH/H$_2$CO column density is here plotted as a function of the H$_2$CO rotational temperature. The semi-transparent data points represent embedded sources in the CrA star-forming region \citep{lindberg15}. Red data points represent the secondary component dominated by H$_2$CO, CH$_3$OH, and SO in sources with two line components (see \autoref{sec:secondary}).}
	\label{fig:ch3oh_h2co}
\end{figure}

Surprisingly, the same differentiation in temperature between H$_2$CO and \ctht\ is found toward IRAS~16293\nobr2422, where the high H$_2$CO temperature must have an internal origin (see above). The \ctht\ gas likely also exists in this hot inner envelope, but the bulk of the emission possibly traces an intermediately deep region of the protostellar envelope, too deeply embedded to be heated by external irradiation, but also far from the internal heating source that gives rise to the \textit{hot corino} chemistry of the inner 50--100~au of IRAS~16293\nobr2422, where the temperature is $>100$~K, making complex organic molecules as well as H$_2$CO evaporate from the icy grain mantles.

\section{Correlations between different species}

In this section, to evaluate the statistical significance of any correlations, we use Spearman's $\rho$ rank correlation test, which tests the monotonicity of two sets of variables without requiring a linear relation or a normal distribution of the variables. The H$_2$CO and \ctht\ rotational temperatures and column densities are given in \autoref{tab:rotdiagresults}. The CH$_3$OH column densities were estimated using the CH$_3$OH line at 218.440~GHz and the H$_2$CO rotational temperature as a proxy for the CH$_3$OH rotational temperature. This assumption is based on the similar origin of these two molecules (hydrogenation of CO on grain mantles; see e.g. \citealt{charnley97}), interferometric observations showing them to be spatially co-aligned in protostellar envelopes \citep[e.g.][]{lindberg12}, and the similar critical densities of the observed H$_2$CO and CH$_3$OH lines (a few~$\times~10^5$~cm$^{-3}$; \citealt{guzman11,guzman13}). We used RADEX models to find that the CH$_3$OH line at 218.440~GHz is optically thin and that our method provides reliable values on the CH$_3$OH column density assuming $n\sim10^6$~cm$^{-3}$ and $T\lesssim40$~K \citep[see also][]{lindberg16}.

\citet{lindberg15} suggested a trend between the H$_2$CO rotational temperature and the CH$_3$OH/H$_2$CO ratio in the embedded protostars of CrA.  In \autoref{fig:ch3oh_h2co} we plot the CH$_3$OH/H$_2$CO column density ratio as a function of the H$_2$CO rotational temperature, and find that they have a strong correlation (Spearman's rank test gives $\rho=0.62$ with $p<0.001$). If disregarding the data points of the \textit{hot corino} source IRAS~16293\nobr2422 (since the bulk of this emission should originate in the hot inner envelope), the correlation weakens somewhat ($\rho=0.52$ with $p<0.01$). However, if also disregarding the secondary components due to their uncertain origin, there is no significant correlation between the CH$_3$OH/H$_2$CO and the H$_2$CO rotational temperature. A correlation would be in agreement with laboratory results suggesting that the grain-surface hydrogenation reactions forming CH$_3$OH are more efficient at higher temperatures \citep[e.g.][]{fuchs09}. The exact mechanism behind CH$_3$OH evaporation at low temperatures, however, remains unclear \citep{martin16,bertin16}.

\begin{figure*}[t]
	\epsscale{1.1}
	\plottwo{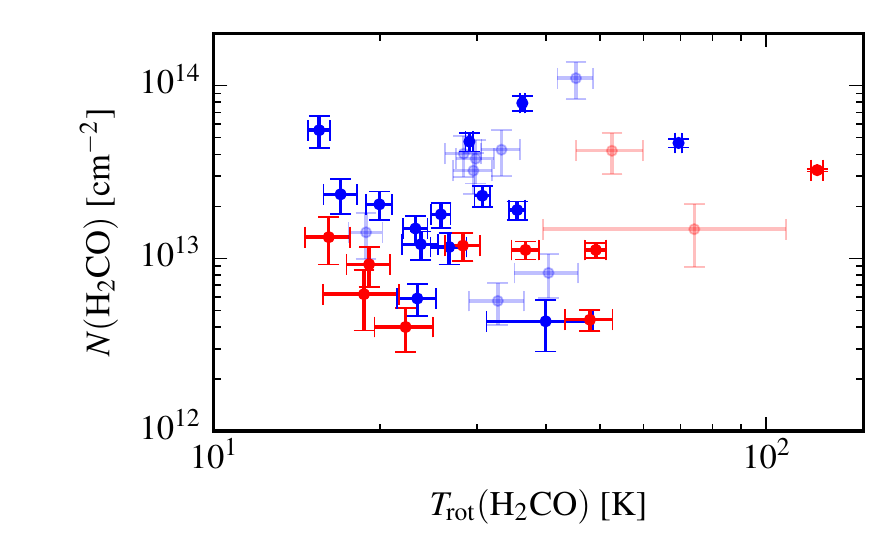}{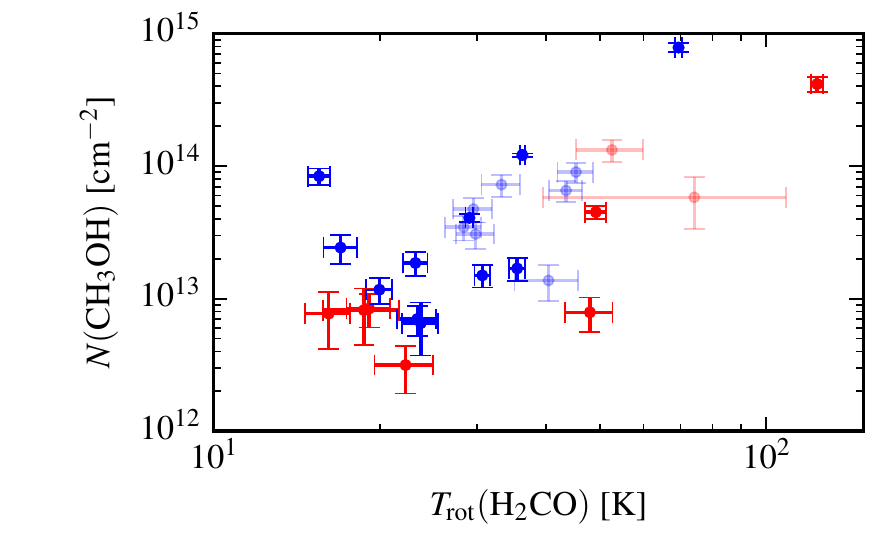}
	\plottwo{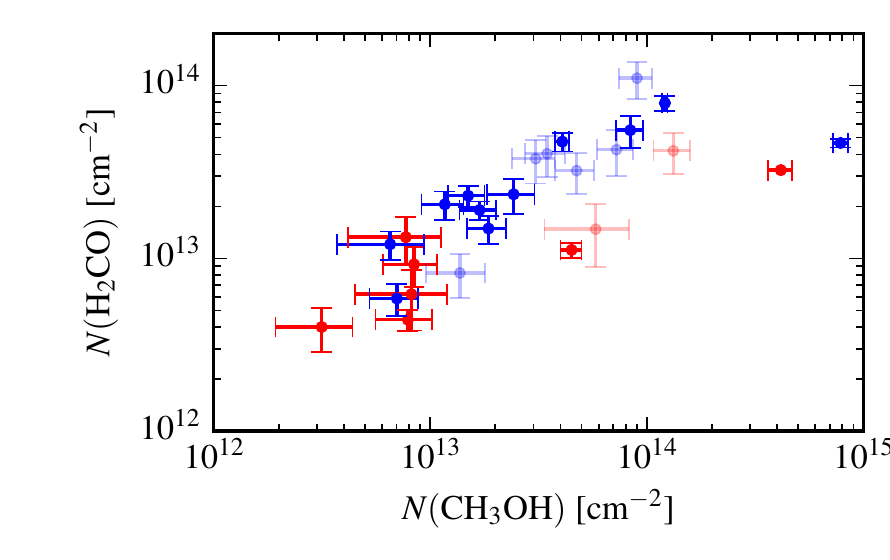}{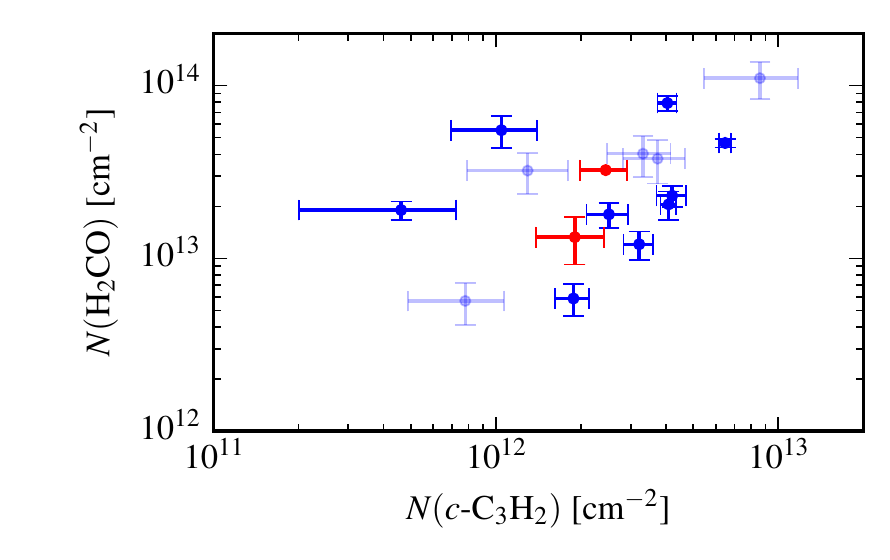}
	\caption{Plots studying trends between rotational temperatures and column densities of H$_2$CO, CH$_3$OH, and \ctht. The CH$_3$OH column density was estimated assuming $T_\mathrm{rot}($CH$_3$OH$) = T_\mathrm{rot}($H$_2$CO). The semi-transparent data points represent similar embedded sources in the CrA star-forming region \citep{lindberg15}. Red data points represent the secondary component dominated by H$_2$CO, CH$_3$OH, and SO in sources with two line components (see \autoref{sec:secondary}).}
	\label{fig:correlations}
\end{figure*}

In \autoref{fig:correlations}, we investigate the correlation between the rotational temperatures and column densities of H$_2$CO, CH$_3$OH, and \ctht. The CH$_3$OH column density was estimated assuming $T_\mathrm{rot}($CH$_3$OH$) = T_\mathrm{rot}($H$_2$CO). We find no correlation between the H$_2$CO rotational temperature and column density ($\rho=0.14$, $p\approx0.44$), but the CH$_3$OH column density is moderately correlated with the H$_2$CO temperature ($\rho=0.58$, $p\approx0.001$), which indicates that CH$_3$OH production is enhanced by high temperatures or radiation fields (grain-mantle evaporation and/or enhancement of hydrogenation reactions on grain surfaces; cf.\ \citealt{oberg09}). If disregarding the \textit{hot corino} source IRAS~16293\nobr2422, the correlation is still significant ($\rho=0.48$, $p<0.02$).
For \ctht\ there is no significant correlation between its rotational temperature and column density (not plotted; $\rho=0.19$, $p\approx0.41$), which would suggest that its formation is not temperature-dependent.
We find a very strong correlation between the H$_2$CO and CH$_3$OH column densities ($\rho=0.82$, $p<10^{-6}$) indicating that these molecules share a common origin (e.g. grain-mantle hydrogenation of CO). To exclude the possibility of this correlation being biased by the fact that we use the H$_2$CO temperature to find the CH$_3$OH column density we also performed this calculation assuming a fixed $T_\mathrm{rot}($CH$_3$OH$) = 10$~K for all sources, and still find a strong correlation between $N($H$_2$CO$)$ and $N($CH$_3$OH$)$. Finally, the \ctht\ and H$_2$CO column densities show a barely significant correlation ($\rho=0.52$, $p\approx0.04$).

\section{Summary and Conclusions}

We find that the H$_2$CO temperature in protostellar cores depends strongly on the distance to external sources of strong irradiation, whereas the \ctht\ temperature is insensitive to such irradiation (see \autoref{h2co_temp}). This implies that the H$_2$CO traces the outer, externally irradiated layers of the protostellar cores, whereas the emission from \ctht\ originates primarily in the inner, more shielded region. However, in the \textit{hot corino} source IRAS~16293\nobr2422, the majority of the \ctht\ gas appears to mainly trace an intermediate region, shielded from any external irradiation, but sufficiently distant from the central protostar to remain relatively cool.

We also find that the column densities of the two molecules H$_2$CO and CH$_3$OH are closely related, suggesting that they share a common origin. Interferometric observations would be necessary to confirm that they spatially co-exist, and also to study the nature of the secondary components found toward many of the studied sources.

The chemical implications of external heating are still not well-understood. A comparative study using interferometric observations of similar protostars within the same star-forming region but with different levels of external irradiation could be used to investigate the origin of the differences in chemical composition and distribution. We suggest future line surveys searching for complex organic molecules and hydrocarbon species, and interferometric mapping of a few key species in the most luminous sources of $\rho$~Oph~A and $\rho$~Oph~B.

\acknowledgments

This research was supported by NASA's Emerging Worlds Program and by an appointment to the NASA Postdoctoral Program at the NASA Goddard Space Flight Center to J.E.L., administered by Universities Space Research Association through a contract with NASA. We thank the anonymous referee for insightful comments and suggestions which helped us improve the manuscript.

\vspace{1cm}
\bibliographystyle{apj}
\bibliography{oph_heating}

\phantom{text}
\clearpage
\begin{appendix}

\section{Observed line parameters}
\label{app:obslines}

\autoref{tab:obslines} lists the detected lines in all observed sources. The errors on integrated intensities are $1\sigma$ rms errors. In a few of the sources we found that considerable C$^{18}$O emission in the off-position made the C$^{18}$O line data unreliable for those sources. Those are indicated by ``Off-em.''\ in their integrated intensity column. We do not expect any of the other lines, which are much fainter (in particular in the diffuse gas where the off-positions were measured), to be affected by off-position emission.

In sources with two components, these are listed as separate sources. LSR velocities and line widths are from Gaussian fits. For the sources with two components, the integrated intensities are from Gaussian fits. In all other sources, they are integrated over the whole spectral line.

\LongTables
\capstartfalse
\begin{deluxetable*}{lllllr}

	\tablecaption{Measured line parameters}

	\tablehead{
		\colhead{\textbf{Source}} & \colhead{Molecule} & \colhead{Transition} & \colhead{$v_\mathrm{LSR}$} & \colhead{$\Delta v$} & \colhead{$\int T_\mathrm{mb}dv$} \\
		\colhead{Rest freq.} \\
		\colhead{[MHz]} & \colhead{} & \colhead{} & \colhead{[km~s$^{-1}$]} & \colhead{[km~s$^{-1}$]} & \colhead{[K~km~s$^{-1}$]}}
	
	%% All data must appear between the \startdata and \enddata commands
	\startdata
	\vspace{-6pt}\\
\multicolumn6l{\textbf{GDS J162625.6 (main)}} \\ 
216112.58 & DCO$^+$ & $J=3$--$2$ & 3.8 & 0.8 & $3.84\pm0.03$ \\ 
216278.76 & $c$-C$_3$H$_2$ & $3_{03}$--$2_{21}$ & 3.8 & 0.8 & $0.28\pm0.02$ \\ 
216373.32 & C$_2$D & $N=3$--$2,J=7/2$--$5/2$\tablenotemark{a} & 4.0 & 1.3 & $0.48\pm0.03$ \\ 
216428.32 & C$_2$D & $N=3$--$2,J=5/2$--$3/2$\tablenotemark{b} & 3.6 & 1.0 & $0.28\pm0.02$ \\ 
217238.54 & DCN & $J=3$--$2$ & 3.8 & 1.1 & $0.58\pm0.03$ \\ 
217822.15 & $c$-C$_3$H$_2$ & $6_{06}$--$5_{15}$, $6_{16}$--$5_{05}$ & 3.8 & 0.8 & $0.38\pm0.03$ \\ 
217940.05 & $c$-C$_3$H$_2$ & $5_{14}$--$4_{23}$ & 3.8 & 0.7 & $0.22\pm0.02$ \\ 
218160.46 & $c$-C$_3$H$_2$ & $5_{24}$--$4_{13}$ & 3.9 & 1.3 & $0.09\pm0.02$ \\ 
218222.19 & H$_2$CO & $3_{03}$--$2_{02}$ & 3.9 & 0.8 & $1.87\pm0.02$ \\ 
218475.63 & H$_2$CO & $3_{22}$--$2_{21}$ & 3.8 & 0.7 & $0.17\pm0.02$ \\ 
218760.07 & H$_2$CO & $3_{21}$--$2_{20}$ & 3.8 & 0.9 & $0.16\pm0.02$ \\ 
219560.35 & C$^{18}$O & $J=2$--$1$ & 3.9 & 0.8 & $10.65\pm0.02$ \\ 
219949.44 & SO & $6_5$--$5_4$ & 4.0 & 0.7 & $1.45\pm0.02$ \\
\vspace{-6pt}\\
\multicolumn6l{\textbf{GDS J162625.6 (secondary)}} \\* 
218222.19 & H$_2$CO & $3_{03}$--$2_{02}$ & 2.7 & 1.4 & $1.56\pm0.02$ \\ 
218475.63 & H$_2$CO & $3_{22}$--$2_{21}$ & 2.6 & 1.7 & $0.28\pm0.02$ \\ 
218760.07 & H$_2$CO & $3_{21}$--$2_{20}$ & 2.5 & 1.3 & $0.18\pm0.02$ \\ 
219560.35 & C$^{18}$O & $J=2$--$1$ & 3.0 & 1.6 & $9.76\pm0.03$ \\ 
219949.44 & SO & $6_5$--$5_4$ & 2.8 & 1.9 & $2.30\pm0.03$ \\ 
\vspace{-6pt}\\
\multicolumn6l{\textbf{GSS 26}} \\ *
218222.19 & H$_2$CO & $3_{03}$--$2_{02}$ & 3.4 & 0.9 & $0.65\pm0.04$ \\ 
218475.63 & H$_2$CO & $3_{22}$--$2_{21}$ & 3.2 & 1.7 & $0.12\pm0.04$ \\ 
218760.07 & H$_2$CO & $3_{21}$--$2_{20}$ & 3.6 & 1.1 & $0.10\pm0.04$ \\ 
219560.35 & C$^{18}$O & $J=2$--$1$ & 3.4 & 1.6 & $19.40\pm0.05$ \\ 
219949.44 & SO & $6_5$--$5_4$ & 3.4 & 1.3 & $0.71\pm0.03$ \\ 
\vspace{-6pt}\\
\multicolumn6l{\textbf{GSS 30}} \\ *
216112.58 & DCO$^+$ & $J=3$--$2$ & 3.2 & 0.7 & $0.27\pm0.02$ \\ 
216278.76 & $c$-C$_3$H$_2$ & $3_{03}$--$2_{21}$ & 3.4 & 0.4 & $0.05\pm0.01$ \\ 
217104.98 & SiO & $J=5$--$4$ & 3.2 & 9.8 & $0.77\pm0.06$ \\ 
217822.15 & $c$-C$_3$H$_2$ & $6_{06}$--$5_{15}$, $6_{16}$--$5_{05}$ & 3.5 & 1.4 & $0.10\pm0.02$ \\ 
217827.18 & $^{33}$SO & $6_5$--$5_4, F=9/2$--$7/2$ & 3.0 & 0.9 & $0.17\pm0.02$ \\ 
217829.83 & $^{33}$SO & $6_5$--$5_4$ & 3.0 & 1.2 & $0.25\pm0.02$ \\ 
217831.77 & $^{33}$SO & $6_5$--$5_4$\tablenotemark{c} & 2.9 & 1.4 & $0.50\pm0.02$ \\ 
218222.19 & H$_2$CO & $3_{03}$--$2_{02}$ & 3.3 & 1.3 & $2.52\pm0.03$ \\ 
218324.72 & HC$_3$N & $J=24$--$23$ & 3.1 & 0.3 & $0.03\pm0.01$ \\ 
218440.06 & CH$_3$OH & $4_{2}$--$3_{1}$,~E  & 3.1 & 1.1 & $0.10\pm0.02$ \\ 
218475.63 & H$_2$CO & $3_{22}$--$2_{21}$ & 3.3 & 1.3 & $0.41\pm0.02$ \\ 
218760.07 & H$_2$CO & $3_{21}$--$2_{20}$ & 3.4 & 1.3 & $0.28\pm0.03$ \\ 
219166.24 & CCO & $N=10$--$9, J=9$--$9$ & 2.3 & 1.1 & $0.22\pm0.02$ \\ 
219355.01 & $^{34}$SO$_2$ & $11_{1,11}$--$10_{0,10}$ & 3.1 & 1.1 & $0.33\pm0.02$ \\ 
219560.35 & C$^{18}$O & $J=2$--$1$ & 3.3 & 1.6 & $16.20\pm0.03$ \\ 
219949.44 & SO & $6_5$--$5_4$ & 3.4 & 1.4 & $15.40\pm0.03$ \\ 
\vspace{-6pt}\\
\multicolumn6l{\textbf{GSS 39}} \\ *
219560.35 & C$^{18}$O & $J=2$--$1$ & ... & ... & Off-em. \\ 
\vspace{-6pt}\\
\multicolumn6l{\textbf{{[}GY92{]} 30}} \\ *
216112.58 & DCO$^+$ & $J=3$--$2$ & 3.4 & 0.6 & $1.03\pm0.02$ \\ 
216278.76 & $c$-C$_3$H$_2$ & $3_{03}$--$2_{21}$ & 3.4 & 0.4 & $0.42\pm0.01$ \\ 
216373.32 & C$_2$D & $N=3$--$2,J=7/2$--$5/2$\tablenotemark{a} & 3.8 & 1.3 & $0.24\pm0.02$ \\ 
216428.32 & C$_2$D & $N=3$--$2,J=5/2$--$3/2$\tablenotemark{b} & 3.2 & 1.1 & $0.14\pm0.02$ \\ 
217238.54 & DCN & $J=3$--$2$ & 3.3 & 1.0 & $0.31\pm0.02$ \\ 
217822.15 & $c$-C$_3$H$_2$ & $6_{06}$--$5_{15}$, $6_{16}$--$5_{05}$ & 3.4 & 0.5 & $0.39\pm0.02$ \\ 
217940.05 & $c$-C$_3$H$_2$ & $5_{14}$--$4_{23}$ & 3.4 & 0.4 & $0.24\pm0.01$ \\ 
218160.46 & $c$-C$_3$H$_2$ & $5_{24}$--$4_{13}$ & 3.4 & 0.3 & $0.08\pm0.01$ \\ 
218222.19 & H$_2$CO & $3_{03}$--$2_{02}$ & 3.4 & 1.3 & $8.01\pm0.03$ \\ 
218440.06 & CH$_3$OH & $4_{2}$--$3_{1}$,~E  & 3.3 & 1.1 & $0.72\pm0.02$ \\ 
218475.63 & H$_2$CO & $3_{22}$--$2_{21}$ & 3.3 & 1.2 & $1.25\pm0.02$ \\ 
218760.07 & H$_2$CO & $3_{21}$--$2_{20}$ & 3.4 & 1.2 & $1.18\pm0.02$ \\ 
219560.35 & C$^{18}$O & $J=2$--$1$ & 3.3 & 1.3 & $19.50\pm0.03$ \\ 
219798.27 & HNCO & $10_{0,10}$--$9_{0,9}$ & 3.2 & 0.8 & $0.07\pm0.01$ \\ 
219908.52 & H$_2^{13}$CO & $3_{12}$--$2_{11}$ & 3.4 & 1.2 & $0.23\pm0.02$ \\ 
219949.44 & SO & $6_5$--$5_4$ & 3.2 & 1.1 & $8.26\pm0.02$ \\ 
\vspace{-6pt}\\
\multicolumn6l{\textbf{{[}GY92{]} 91}} \\ *
218222.19 & H$_2$CO & $3_{03}$--$2_{02}$ & 3.2 & 0.7 & $1.29\pm0.03$ \\ 
218475.63 & H$_2$CO & $3_{22}$--$2_{21}$ & 3.1 & 0.6 & $0.10\pm0.02$ \\ 
218760.07 & H$_2$CO & $3_{21}$--$2_{20}$ & 3.2 & 0.6 & $0.14\pm0.02$ \\ 
219560.35 & C$^{18}$O & $J=2$--$1$ & ... & ... & Off-em. \\ 
219949.44 & SO & $6_5$--$5_4$ & 3.2 & 0.8 & $0.64\pm0.02$ \\ 
\vspace{-6pt}\\
\multicolumn6l{\textbf{ISO-Oph 21 (main)}} \\ *
216112.58 & DCO$^+$ & $J=3$--$2$ & 2.5 & 0.8 & $0.12\pm0.02$ \\ 
218222.19 & H$_2$CO & $3_{03}$--$2_{02}$ & 2.8 & 1.0 & $1.50\pm0.02$ \\ 
218440.06 & CH$_3$OH & $4_{2}$--$3_{1}$,~E  & 2.8 & 1.6 & $0.12\pm0.02$ \\ 
218475.63 & H$_2$CO & $3_{22}$--$2_{21}$ & 2.8 & 0.9 & $0.13\pm0.02$ \\ 
218760.07 & H$_2$CO & $3_{21}$--$2_{20}$ & 2.9 & 0.8 & $0.08\pm0.02$ \\ 
219560.35 & C$^{18}$O & $J=2$--$1$ & 2.8 & 0.8 & $9.29\pm0.02$ \\ 
219949.44 & SO & $6_5$--$5_4$ & 2.8 & 0.8 & $2.89\pm0.02$ \\ 
\vspace{-6pt}\\
\multicolumn6l{\textbf{ISO-Oph 21 (secondary)}} \\ *
218222.19 & H$_2$CO & $3_{03}$--$2_{02}$ & 3.9 & 0.9 & $0.71\pm0.02$ \\ 
218440.06 & CH$_3$OH & $4_{2}$--$3_{1}$,~E  & 4.1 & 0.4 & $0.04\pm0.01$ \\ 
218475.63 & H$_2$CO & $3_{22}$--$2_{21}$ & 4.0 & 1.2 & $0.16\pm0.02$ \\ 
218760.07 & H$_2$CO & $3_{21}$--$2_{20}$ & 3.9 & 1.4 & $0.14\pm0.02$ \\ 
219560.35 & C$^{18}$O & $J=2$--$1$ & 4.1 & 1.0 & $5.59\pm0.02$ \\ 
219949.44 & SO & $6_5$--$5_4$ & 3.8 & 1.2 & $0.96\pm0.02$ \\ 
\vspace{-6pt}\\
\multicolumn6l{\textbf{J162614.6 (main)}} \\ *
216112.58 & DCO$^+$ & $J=3$--$2$ & 3.0 & 0.7 & $0.09\pm0.02$ \\ 
218222.19 & H$_2$CO & $3_{03}$--$2_{02}$ & 2.7 & 1.0 & $4.45\pm0.02$ \\ 
218440.06 & CH$_3$OH & $4_{2}$--$3_{1}$,~E  & 2.8 & 1.0 & $0.26\pm0.02$ \\ 
218475.63 & H$_2$CO & $3_{22}$--$2_{21}$ & 2.7 & 0.9 & $0.54\pm0.02$ \\ 
218760.07 & H$_2$CO & $3_{21}$--$2_{20}$ & 2.7 & 0.8 & $0.42\pm0.02$ \\ 
219560.35 & C$^{18}$O & $J=2$--$1$ & 2.7 & 0.9 & $10.95\pm0.02$ \\ 
219908.52 & H$_2^{13}$CO & $3_{12}$--$2_{11}$ & 2.9 & 0.7 & $0.09\pm0.02$ \\ 
219949.44 & SO & $6_5$--$5_4$ & 2.8 & 0.9 & $4.14\pm0.02$ \\ 
\vspace{-6pt}\\
\multicolumn6l{\textbf{J162614.6 (secondary)}} \\ *
218222.19 & H$_2$CO & $3_{03}$--$2_{02}$ & 4.4 & 1.5 & $1.79\pm0.03$ \\ 
218440.06 & CH$_3$OH & $4_{2}$--$3_{1}$,~E  & 4.3 & 1.6 & $0.22\pm0.02$ \\ 
218475.63 & H$_2$CO & $3_{22}$--$2_{21}$ & 4.3 & 1.3 & $0.36\pm0.02$ \\ 
218760.07 & H$_2$CO & $3_{21}$--$2_{20}$ & 4.2 & 1.5 & $0.41\pm0.02$ \\ 
219560.35 & C$^{18}$O & $J=2$--$1$ & 4.6 & 1.0 & $2.04\pm0.02$ \\ 
219908.52 & H$_2^{13}$CO & $3_{12}$--$2_{11}$ & 4.0 & 1.1 & $0.10\pm0.02$ \\ 
219949.44 & SO & $6_5$--$5_4$ & 4.0 & 1.9 & $0.89\pm0.03$ \\ 
\vspace{-6pt}\\
\multicolumn6l{\textbf{VLA 1623 (main)}} \\ *
216112.58 & DCO$^+$ & $J=3$--$2$ & 3.8 & 0.8 & $4.85\pm0.03$ \\ 
216278.76 & $c$-C$_3$H$_2$ & $3_{03}$--$2_{21}$ & 3.8 & 0.8 & $0.45\pm0.03$ \\ 
216373.32 & C$_2$D & $N=3$--$2,J=7/2$--$5/2$\tablenotemark{a} & 4.0 & 1.2 & $0.64\pm0.03$ \\ 
216428.32 & C$_2$D & $N=3$--$2,J=5/2$--$3/2$\tablenotemark{b} & 3.6 & 0.8 & $0.41\pm0.02$ \\ 
217238.54 & DCN & $J=3$--$2$ & 3.8 & 0.8 & $0.55\pm0.03$ \\ 
217822.15 & $c$-C$_3$H$_2$ & $6_{06}$--$5_{15}$, $6_{16}$--$5_{05}$ & 3.8 & 0.7 & $0.52\pm0.02$ \\ 
217940.05 & $c$-C$_3$H$_2$ & $5_{14}$--$4_{23}$ & 3.8 & 0.7 & $0.34\pm0.02$ \\ 
218160.46 & $c$-C$_3$H$_2$ & $5_{24}$--$4_{13}$ & 3.8 & 0.9 & $0.09\pm0.03$ \\ 
218222.19 & H$_2$CO & $3_{03}$--$2_{02}$ & 3.8 & 0.8 & $2.57\pm0.02$ \\ 
218440.06 & CH$_3$OH & $4_{2}$--$3_{1}$,~E  & 3.7 & 0.5 & $0.09\pm0.02$ \\ 
218475.63 & H$_2$CO & $3_{22}$--$2_{21}$ & 3.7 & 0.8 & $0.33\pm0.02$ \\ 
218760.07 & H$_2$CO & $3_{21}$--$2_{20}$ & 3.8 & 0.8 & $0.27\pm0.02$ \\ 
219560.35 & C$^{18}$O & $J=2$--$1$ & 3.9 & 0.8 & $13.10\pm0.02$ \\ 
219908.52 & H$_2^{13}$CO & $3_{12}$--$2_{11}$ & 3.8 & 0.7 & $0.10\pm0.02$ \\ 
219949.44 & SO & $6_5$--$5_4$ & 3.9 & 0.7 & $2.98\pm0.02$ \\ 
\vspace{-6pt}\\
\multicolumn6l{\textbf{VLA 1623 (secondary)}} \\ *
218222.19 & H$_2$CO & $3_{03}$--$2_{02}$ & 2.5 & 1.6 & $1.46\pm0.03$ \\ 
218475.63 & H$_2$CO & $3_{22}$--$2_{21}$ & 2.3 & 1.3 & $0.16\pm0.02$ \\ 
218760.07 & H$_2$CO & $3_{21}$--$2_{20}$ & 2.8 & 1.4 & $0.14\pm0.03$ \\ 
219560.35 & C$^{18}$O & $J=2$--$1$ & 2.6 & 1.5 & $6.26\pm0.03$ \\ 
219949.44 & SO & $6_5$--$5_4$ & 2.3 & 1.5 & $1.96\pm0.03$ \\ 
\vspace{-6pt}\\
\multicolumn6l{\textbf{ISO-Oph 124 (main)}} \\ *
216112.58 & DCO$^+$ & $J=3$--$2$ & 3.9 & 0.4 & $0.24\pm0.01$ \\ 
218222.19 & H$_2$CO & $3_{03}$--$2_{02}$ & 4.0 & 1.1 & $1.71\pm0.02$ \\ 
218440.06 & CH$_3$OH & $4_{2}$--$3_{1}$,~E  & 3.9 & 1.4 & $0.13\pm0.02$ \\ 
218475.63 & H$_2$CO & $3_{22}$--$2_{21}$ & 4.7 & 1.4 & $0.08\pm0.02$ \\ 
218760.07 & H$_2$CO & $3_{21}$--$2_{20}$ & 4.3 & 0.5 & $0.04\pm0.01$ \\ 
219560.35 & C$^{18}$O & $J=2$--$1$ & 4.2 & 1.2 & $12.00\pm0.02$ \\ 
219949.44 & SO & $6_5$--$5_4$ & 4.1 & 1.0 & $0.81\pm0.02$ \\ 
\vspace{-6pt}\\
\multicolumn6l{\textbf{ISO-Oph 124 (secondary)}} \\ *
216112.58 & DCO$^+$ & $J=3$--$2$ & 3.1 & 0.4 & $0.10\pm0.01$ \\ 
218222.19 & H$_2$CO & $3_{03}$--$2_{02}$ & 3.0 & 0.5 & $0.76\pm0.01$ \\ 
218440.06 & CH$_3$OH & $4_{2}$--$3_{1}$,~E  & 3.1 & 0.2 & $0.05\pm0.01$ \\ 
218475.63 & H$_2$CO & $3_{22}$--$2_{21}$ & 3.2 & 0.5 & $0.04\pm0.01$ \\ 
218760.07 & H$_2$CO & $3_{21}$--$2_{20}$ & 3.1 & 0.3 & $0.03\pm0.01$ \\ 
219560.35 & C$^{18}$O & $J=2$--$1$ & 3.1 & 0.7 & $3.90\pm0.02$ \\ 
219949.44 & SO & $6_5$--$5_4$ & 3.1 & 0.4 & $0.35\pm0.01$ \\ 
\vspace{-6pt}\\
\multicolumn6l{\textbf{J162728 (main)}} \\ *
216112.58 & DCO$^+$ & $J=3$--$2$ & 4.2 & 0.5 & $2.69\pm0.01$ \\ 
216278.76 & $c$-C$_3$H$_2$ & $3_{03}$--$2_{21}$ & 4.3 & 0.4 & $0.43\pm0.01$ \\ 
216373.32 & C$_2$D & $N=3$--$2,J=7/2$--$5/2$\tablenotemark{a} & 5.0 & 0.3 & $0.07\pm0.01$ \\ 
216428.32 & C$_2$D & $N=3$--$2,J=5/2$--$3/2$\tablenotemark{b} & 4.3 & 0.5 & $0.11\pm0.01$ \\ 
217238.54 & DCN & $J=3$--$2$ & 4.2 & 0.5 & $0.12\pm0.02$ \\ 
217822.15 & $c$-C$_3$H$_2$ & $6_{06}$--$5_{15}$, $6_{16}$--$5_{05}$ & 4.3 & 0.4 & $0.41\pm0.01$ \\ 
217940.05 & $c$-C$_3$H$_2$ & $5_{14}$--$4_{23}$ & 4.3 & 0.5 & $0.27\pm0.01$ \\ 
218160.46 & $c$-C$_3$H$_2$ & $5_{24}$--$4_{13}$ & 4.3 & 0.4 & $0.10\pm0.01$ \\ 
218222.19 & H$_2$CO & $3_{03}$--$2_{02}$ & 4.3 & 0.6 & $1.58\pm0.01$ \\ 
218440.06 & CH$_3$OH & $4_{2}$--$3_{1}$,~E  & 4.2 & 0.6 & $0.07\pm0.01$ \\ 
218475.63 & H$_2$CO & $3_{22}$--$2_{21}$ & 4.2 & 0.7 & $0.09\pm0.01$ \\ 
218760.07 & H$_2$CO & $3_{21}$--$2_{20}$ & 4.2 & 0.6 & $0.06\pm0.01$ \\ 
219560.35 & C$^{18}$O & $J=2$--$1$ & 4.3 & 0.6 & $4.43\pm0.02$ \\ 
219949.44 & SO & $6_5$--$5_4$ & 4.2 & 0.7 & $0.49\pm0.02$ \\ 
\vspace{-6pt}\\
\multicolumn6l{\textbf{J162728 (secondary)}} \\ *
216112.58 & DCO$^+$ & $J=3$--$2$ & 3.4 & 0.5 & $0.92\pm0.02$ \\ 
216278.76 & $c$-C$_3$H$_2$ & $3_{03}$--$2_{21}$ & 3.9 & 0.5 & $0.14\pm0.01$ \\ 
216373.32 & C$_2$D & $N=3$--$2,J=7/2$--$5/2$\tablenotemark{a} & 4.3 & 0.6 & $0.17\pm0.01$ \\ 
216428.32 & C$_2$D & $N=3$--$2,J=5/2$--$3/2$\tablenotemark{b} & 3.4 & 0.5 & $0.04\pm0.01$ \\ 
217822.15 & $c$-C$_3$H$_2$ & $6_{06}$--$5_{15}$, $6_{16}$--$5_{05}$ & 3.7 & 0.5 & $0.07\pm0.01$ \\ 
217940.05 & $c$-C$_3$H$_2$ & $5_{14}$--$4_{23}$ & 3.2 & 0.1 & $0.01\pm0.01$ \\ 
218222.19 & H$_2$CO & $3_{03}$--$2_{02}$ & 3.4 & 0.7 & $0.93\pm0.02$ \\ 
218440.06 & CH$_3$OH & $4_{2}$--$3_{1}$,~E  & 3.2 & 0.7 & $0.04\pm0.02$ \\ 
218475.63 & H$_2$CO & $3_{22}$--$2_{21}$ & 3.3 & 0.5 & $0.03\pm0.01$ \\ 
218760.07 & H$_2$CO & $3_{21}$--$2_{20}$ & 3.5 & 0.2 & $0.02\pm0.01$ \\ 
219560.35 & C$^{18}$O & $J=2$--$1$ & 3.1 & 1.3 & $8.05\pm0.02$ \\ 
219949.44 & SO & $6_5$--$5_4$ & 3.4 & 0.5 & $0.40\pm0.01$ \\ 
\vspace{-6pt}\\
\multicolumn6l{\textbf{Oph-emb 5}} \\ *
216112.58 & DCO$^+$ & $J=3$--$2$ & 3.9 & 0.8 & $2.46\pm0.02$ \\ 
216278.76 & $c$-C$_3$H$_2$ & $3_{03}$--$2_{21}$ & 3.9 & 0.5 & $0.10\pm0.01$ \\ 
217822.15 & $c$-C$_3$H$_2$ & $6_{06}$--$5_{15}$, $6_{16}$--$5_{05}$ & 4.0 & 1.0 & $0.08\pm0.02$ \\ 
217940.05 & $c$-C$_3$H$_2$ & $5_{14}$--$4_{23}$ & 3.7 & 0.4 & $0.04\pm0.01$ \\ 
218222.19 & H$_2$CO & $3_{03}$--$2_{02}$ & 3.9 & 1.4 & $3.17\pm0.02$ \\ 
218440.06 & CH$_3$OH & $4_{2}$--$3_{1}$,~E  & 3.8 & 1.0 & $0.42\pm0.02$ \\ 
218475.63 & H$_2$CO & $3_{22}$--$2_{21}$ & 3.7 & 0.8 & $0.09\pm0.02$ \\ 
218760.07 & H$_2$CO & $3_{21}$--$2_{20}$ & 3.7 & 0.7 & $0.08\pm0.02$ \\ 
219560.35 & C$^{18}$O & $J=2$--$1$ & 3.7 & 1.3 & $9.43\pm0.03$ \\ 
219908.52 & H$_2^{13}$CO & $3_{12}$--$2_{11}$ & 3.8 & 0.6 & $0.07\pm0.02$ \\ 
219949.44 & SO & $6_5$--$5_4$ & 3.8 & 1.1 & $3.27\pm0.02$ \\ 
\vspace{-6pt}\\
\multicolumn6l{\textbf{VSSG 17 (main)}} \\ *
216112.58 & DCO$^+$ & $J=3$--$2$ & 4.2 & 0.6 & $1.12\pm0.03$ \\ 
216278.76 & $c$-C$_3$H$_2$ & $3_{03}$--$2_{21}$ & 4.3 & 0.6 & $0.35\pm0.02$ \\ 
216373.32 & C$_2$D & $N=3$--$2,J=7/2$--$5/2$\tablenotemark{a} & 4.4 & 1.2 & $0.12\pm0.02$ \\ 
216428.32 & C$_2$D & $N=3$--$2,J=5/2$--$3/2$\tablenotemark{b} & 4.2 & 0.4 & $0.06\pm0.02$ \\ 
217238.54 & DCN & $J=3$--$2$ & 4.4 & 0.6 & $0.06\pm0.02$ \\ 
217822.15 & $c$-C$_3$H$_2$ & $6_{06}$--$5_{15}$, $6_{16}$--$5_{05}$ & 4.3 & 0.5 & $0.40\pm0.02$ \\ 
217940.05 & $c$-C$_3$H$_2$ & $5_{14}$--$4_{23}$ & 4.3 & 0.4 & $0.24\pm0.02$ \\ 
218160.46 & $c$-C$_3$H$_2$ & $5_{24}$--$4_{13}$ & 4.4 & 0.6 & $0.05\pm0.02$ \\ 
218222.19 & H$_2$CO & $3_{03}$--$2_{02}$ & 4.4 & 0.8 & $1.22\pm0.02$ \\ 
218440.06 & CH$_3$OH & $4_{2}$--$3_{1}$,~E  & 4.5 & 0.8 & $0.04\pm0.02$ \\ 
218475.63 & H$_2$CO & $3_{22}$--$2_{21}$ & 4.4 & 1.1 & $0.11\pm0.02$ \\ 
218760.07 & H$_2$CO & $3_{21}$--$2_{20}$ & 4.4 & 0.6 & $0.06\pm0.02$ \\ 
219560.35 & C$^{18}$O & $J=2$--$1$ & 4.5 & 1.1 & $5.80\pm0.02$ \\ 
219949.44 & SO & $6_5$--$5_4$ & 4.5 & 0.7 & $0.18\pm0.02$ \\ 
\vspace{-6pt}\\
\multicolumn6l{\textbf{VSSG 17 (secondary)}} \\ *
218222.19 & H$_2$CO & $3_{03}$--$2_{02}$ & 3.3 & 0.8 & $0.55\pm0.02$ \\ 
218440.06 & CH$_3$OH & $4_{2}$--$3_{1}$,~E  & 3.2 & 0.5 & $0.05\pm0.01$ \\ 
218475.63 & H$_2$CO & $3_{22}$--$2_{21}$ & 3.0 & 0.4 & $0.02\pm0.01$ \\ 
218760.07 & H$_2$CO & $3_{21}$--$2_{20}$ & 3.5 & 0.5 & $0.03\pm0.02$ \\ 
219560.35 & C$^{18}$O & $J=2$--$1$ & 3.1 & 1.2 & $7.16\pm0.02$ \\ 
219949.44 & SO & $6_5$--$5_4$ & 3.4 & 0.7 & $0.11\pm0.01$ \\ 
\vspace{-6pt}\\
\multicolumn6l{\textbf{WL 2}} \\ *
217104.98 & SiO & $J=5$--$4$ & 6.5 & 2.2 & $0.20\pm0.05$ \\ 
218222.19 & H$_2$CO & $3_{03}$--$2_{02}$ & 3.7 & 0.5 & $0.30\pm0.04$ \\ 
219560.35 & C$^{18}$O & $J=2$--$1$ & 3.5 & 1.3 & $7.87\pm0.04$ \\ 
219949.44 & SO & $6_5$--$5_4$ & 3.5 & 2.2 & $0.32\pm0.05$ \\ 
\vspace{-6pt}\\
\multicolumn6l{\textbf{WL 12}} \\ *
216112.58 & DCO$^+$ & $J=3$--$2$ & 4.1 & 0.4 & $0.13\pm0.03$ \\ 
218222.19 & H$_2$CO & $3_{03}$--$2_{02}$ & 4.0 & 0.9 & $0.28\pm0.04$ \\ 
219560.35 & C$^{18}$O & $J=2$--$1$ & 3.9 & 1.2 & $10.10\pm0.05$ \\ 
219949.44 & SO & $6_5$--$5_4$ & 4.1 & 2.8 & $0.43\pm0.06$ \\ 
\vspace{-6pt}\\
\multicolumn6l{\textbf{WL 22}} \\ *
216112.58 & DCO$^+$ & $J=3$--$2$ & 3.7 & 0.6 & $0.62\pm0.02$ \\ 
216278.76 & $c$-C$_3$H$_2$ & $3_{03}$--$2_{21}$ & 3.7 & 0.5 & $0.21\pm0.02$ \\ 
216373.32 & C$_2$D & $N=3$--$2,J=7/2$--$5/2$\tablenotemark{a} & 3.7 & 1.2 & $0.12\pm0.02$ \\ 
217238.54 & DCN & $J=3$--$2$ & 3.8 & 0.7 & $0.07\pm0.02$ \\ 
217822.15 & $c$-C$_3$H$_2$ & $6_{06}$--$5_{15}$, $6_{16}$--$5_{05}$ & 3.8 & 0.4 & $0.10\pm0.01$ \\ 
217940.05 & $c$-C$_3$H$_2$ & $5_{14}$--$4_{23}$ & 3.8 & 0.2 & $0.07\pm0.01$ \\ 
218160.46 & $c$-C$_3$H$_2$ & $5_{24}$--$4_{13}$ & 4.1 & 0.8 & $0.08\pm0.02$ \\ 
218222.19 & H$_2$CO & $3_{03}$--$2_{02}$ & 3.8 & 0.8 & $0.75\pm0.03$ \\ 
218440.06 & CH$_3$OH & $4_{2}$--$3_{1}$,~E  & 3.9 & 0.6 & $0.06\pm0.02$ \\ 
219560.35 & C$^{18}$O & $J=2$--$1$ & ... & ... & Off-em. \\ 
219949.44 & SO & $6_5$--$5_4$ & 3.9 & 0.7 & $0.45\pm0.01$ \\ 
\vspace{-6pt}\\
\multicolumn6l{\textbf{{[}GY92{]} 197}} \\ *
216112.58 & DCO$^+$ & $J=3$--$2$ & 4.8 & 0.3 & $0.08\pm0.02$ \\ 
218222.19 & H$_2$CO & $3_{03}$--$2_{02}$ & 4.8 & 0.9 & $0.34\pm0.03$ \\ 
219560.35 & C$^{18}$O & $J=2$--$1$ & ... & ... & Off-em. \\ 
219949.44 & SO & $6_5$--$5_4$ & 4.8 & 2.8 & $0.29\pm0.04$ \\ 
\vspace{-6pt}\\
\multicolumn6l{\textbf{WL 15}} \\ *
216112.58 & DCO$^+$ & $J=3$--$2$ & 4.6 & 0.6 & $0.10\pm0.03$ \\ 
218222.19 & H$_2$CO & $3_{03}$--$2_{02}$ & 4.5 & 2.2 & $0.44\pm0.05$ \\ 
219560.35 & C$^{18}$O & $J=2$--$1$ & ... & ... & Off-em. \\ 
219949.44 & SO & $6_5$--$5_4$ & 4.9 & 2.3 & $1.78\pm0.06$ \\ 
\vspace{-6pt}\\
\multicolumn6l{\textbf{WL 16}} \\ *
219560.35 & C$^{18}$O & $J=2$--$1$ & ... & ... & Off-em. \\ 
\vspace{-6pt}\\
\multicolumn6l{\textbf{WL 17}} \\ *
218222.19 & H$_2$CO & $3_{03}$--$2_{02}$ & 4.6 & 0.9 & $0.23\pm0.04$ \\ 
219560.35 & C$^{18}$O & $J=2$--$1$ & ... & ... & Off-em. \\ 
219949.44 & SO & $6_5$--$5_4$ & 4.6 & 0.4 & $0.09\pm0.03$ \\ 
\vspace{-6pt}\\
\multicolumn6l{\textbf{IRAS 16244-2432}} \\ *
218222.19 & H$_2$CO & $3_{03}$--$2_{02}$ & 3.7 & 1.9 & $0.51\pm0.05$ \\ 
219560.35 & C$^{18}$O & $J=2$--$1$ & ... & ... & Off-em. \\ 
219949.44 & SO & $6_5$--$5_4$ & 3.8 & 4.2 & $1.83\pm0.07$ \\ 
\vspace{-6pt}\\
\multicolumn6l{\textbf{IRAS 16246-2436}} \\ *
216112.58 & DCO$^+$ & $J=3$--$2$ & 3.6 & 0.4 & $0.50\pm0.02$ \\ 
216278.76 & $c$-C$_3$H$_2$ & $3_{03}$--$2_{21}$ & 3.7 & 0.3 & $0.07\pm0.01$ \\ 
218222.19 & H$_2$CO & $3_{03}$--$2_{02}$ & 3.7 & 0.6 & $0.53\pm0.02$ \\ 
218440.06 & CH$_3$OH & $4_{2}$--$3_{1}$,~E  & 3.7 & 0.6 & $0.09\pm0.01$ \\ 
219560.35 & C$^{18}$O & $J=2$--$1$ & 3.7 & 0.7 & $3.86\pm0.02$ \\ 
219949.44 & SO & $6_5$--$5_4$ & 3.7 & 0.4 & $0.29\pm0.01$ \\ 
\vspace{-6pt}\\
\multicolumn6l{\textbf{ISO-Oph 132}} \\ *
216112.58 & DCO$^+$ & $J=3$--$2$ & 4.0 & 0.3 & $0.26\pm0.03$ \\ 
218222.19 & H$_2$CO & $3_{03}$--$2_{02}$ & 4.0 & 0.3 & $0.18\pm0.02$ \\ 
219560.35 & C$^{18}$O & $J=2$--$1$ & ... & ... & Off-em. \\ 
\vspace{-6pt}\\
\multicolumn6l{\textbf{ISO-Oph 137}} \\ *
216112.58 & DCO$^+$ & $J=3$--$2$ & 4.0 & 0.4 & $1.54\pm0.04$ \\ 
216278.76 & $c$-C$_3$H$_2$ & $3_{03}$--$2_{21}$ & 4.0 & 0.2 & $0.18\pm0.02$ \\ 
216373.32 & C$_2$D & $N=3$--$2,J=7/2$--$5/2$\tablenotemark{a} & 4.6 & 1.2 & $0.21\pm0.04$ \\ 
217238.54 & DCN & $J=3$--$2$ & 4.1 & 0.5 & $0.18\pm0.03$ \\ 
217822.15 & $c$-C$_3$H$_2$ & $6_{06}$--$5_{15}$, $6_{16}$--$5_{05}$ & 3.9 & 0.2 & $0.07\pm0.02$ \\ 
218222.19 & H$_2$CO & $3_{03}$--$2_{02}$ & 4.4 & 1.3 & $0.94\pm0.04$ \\ 
218440.06 & CH$_3$OH & $4_{2}$--$3_{1}$,~E  & 4.7 & 0.2 & $0.05\pm0.02$ \\ 
219560.35 & C$^{18}$O & $J=2$--$1$ & 4.3 & 1.6 & $10.50\pm0.05$ \\ 
219949.44 & SO & $6_5$--$5_4$ & 4.8 & 0.5 & $0.07\pm0.02$ \\ 
\vspace{-6pt}\\
\multicolumn6l{\textbf{ISO-Oph 161}} \\ *
216112.58 & DCO$^+$ & $J=3$--$2$ & 3.6 & 0.4 & $0.95\pm0.03$ \\ 
216278.76 & $c$-C$_3$H$_2$ & $3_{03}$--$2_{21}$ & 3.7 & 0.4 & $0.13\pm0.03$ \\ 
218222.19 & H$_2$CO & $3_{03}$--$2_{02}$ & 3.8 & 0.6 & $0.59\pm0.04$ \\ 
219560.35 & C$^{18}$O & $J=2$--$1$ & ... & ... & Off-em. \\ 
219949.44 & SO & $6_5$--$5_4$ & 3.8 & 0.5 & $0.20\pm0.03$ \\ 
\vspace{-6pt}\\
\multicolumn6l{\textbf{YLW 15}} \\ *
216112.58 & DCO$^+$ & $J=3$--$2$ & 4.1 & 0.4 & $0.87\pm0.03$ \\ 
216278.76 & $c$-C$_3$H$_2$ & $3_{03}$--$2_{21}$ & 4.1 & 0.4 & $0.23\pm0.03$ \\ 
217238.54 & DCN & $J=3$--$2$ & 4.3 & 1.0 & $0.10\pm0.03$ \\ 
217822.15 & $c$-C$_3$H$_2$ & $6_{06}$--$5_{15}$, $6_{16}$--$5_{05}$ & 4.1 & 0.4 & $0.14\pm0.02$ \\ 
218222.19 & H$_2$CO & $3_{03}$--$2_{02}$ & 3.9 & 2.4 & $1.12\pm0.05$ \\ 
219560.35 & C$^{18}$O & $J=2$--$1$ & 4.1 & 1.9 & $11.20\pm0.06$ \\ 
\vspace{-6pt}\\
\multicolumn6l{\textbf{IRAS 16293-2422 (main)}} \\ *
216112.58 & DCO$^+$ & $J=3$--$2$ & 4.5 & 1.1 & $3.62\pm0.02$ \\ 
216278.76 & $c$-C$_3$H$_2$ & $3_{03}$--$2_{21}$ & 4.6 & 1.3 & $0.75\pm0.02$ \\ 
216373.32 & C$_2$D & $N=3$--$2,J=7/2$--$5/2$\tablenotemark{a} & 4.4 & 2.2 & $0.67\pm0.02$ \\ 
216428.32 & C$_2$D & $N=3$--$2,J=5/2$--$3/2$\tablenotemark{b} & 4.0 & 1.8 & $0.33\pm0.02$ \\ 
216570.33 & $^{13}$CN & $N=2$--$1, J=3/2$--$3/2$\tablenotemark{d} & 4.9 & 1.3 & $0.28\pm0.03$ \\ 
216643.30 & SO$_2$ & $22_{2,20}$--$22_{1,21}$ & 5.5 & 2.9 & $0.31\pm0.03$ \\ 
216662.43 & HDCS & $7_{07}$--$6_{06}$ & 4.0 & 1.9 & $0.14\pm0.02$ \\ 
216945.56 & CH$_3$OH & $5_{14}$--$4_{22}$  & 4.6 & 3.4 & $0.53\pm0.03$ \\ 
217104.98 & SiO & $J=5$--$4$ & 4.8 & 5.2 & $5.10\pm0.05$ \\ 
217238.54 & DCN & $J=3$--$2$ & 5.0 & 1.4 & $0.95\pm0.02$ \\ 
217301.18 & $^{13}$CN & $N=2$--$1, J=3/2$--$1/2$\tablenotemark{e} & 4.0 & 6.9 & $0.62\pm0.05$ \\ 
217428.56 & $^{13}$CN & $N=2$--$1, J=5/2$--$3/2$\tablenotemark{f} & 4.5 & 5.9 & $0.25\pm0.04$ \\ 
217469.15 & $^{13}$CN & $N=2$--$1, J=5/2$--$3/2$\tablenotemark{g} & 2.1 & 2.7 & $0.28\pm0.03$ \\ 
217822.15 & $c$-C$_3$H$_2$ & $6_{06}$--$5_{15}$, $6_{16}$--$5_{05}$ & 4.6 & 1.3 & $1.26\pm0.02$ \\ 
217886.39 & CH$_3$OH & $20_1$--$20_0$,~E1 & 3.3 & 4.1 & $0.19\pm0.03$ \\ 
217940.05 & $c$-C$_3$H$_2$ & $5_{14}$--$4_{23}$ & 4.7 & 1.1 & $0.57\pm0.02$ \\ 
218160.46 & $c$-C$_3$H$_2$ & $5_{24}$--$4_{13}$ & 4.7 & 1.9 & $0.35\pm0.02$ \\ 
218222.19 & H$_2$CO & $3_{03}$--$2_{02}$ & 4.8 & 2.0 & $7.37\pm0.02$ \\ 
218324.72 & HC$_3$N & $J=24$--$23$ & 3.9 & 3.7 & $0.62\pm0.03$ \\ 
218440.06 & CH$_3$OH & $4_{2}$--$3_{1}$,~E  & 4.1 & 3.5 & $2.62\pm0.03$ \\ 
218475.63 & H$_2$CO & $3_{22}$--$2_{21}$ & 4.7 & 2.7 & $2.36\pm0.02$ \\ 
218732.73 & $c$-C$_3$H$_2$ & $7_{16}$--$7_{07}$, $7_{26}$--$7_{17}$ & 4.6 & 3.8 & $0.20\pm0.03$ \\ 
218760.07 & H$_2$CO & $3_{21}$--$2_{20}$ & 5.1 & 2.8 & $1.43\pm0.02$ \\ 
218903.36 & OCS & $18$--$17$ & 3.5 & 5.7 & $2.12\pm0.04$ \\ 
219355.01 & $^{34}$SO$_2$ & $11_{1,11}$--$10_{0,10}$ & 4.0 & 2.7 & $0.18\pm0.03$ \\ 
219560.35 & C$^{18}$O & $J=2$--$1$ & 4.5 & 1.6 & $12.46\pm0.02$ \\ 
219908.52 & H$_2^{13}$CO & $3_{12}$--$2_{11}$ & 5.0 & 1.5 & $0.13\pm0.02$ \\ 
219949.44 & SO & $6_5$--$5_4$ & 4.6 & 4.3 & $16.63\pm0.03$ \\ 
\vspace{-6pt}\\
\multicolumn6l{\textbf{IRAS 16293-2422 (secondary)}} \\ *
216112.58 & DCO$^+$ & $J=3$--$2$ & 3.3 & 1.2 & $0.85\pm0.02$ \\ 
216278.76 & $c$-C$_3$H$_2$ & $3_{03}$--$2_{21}$ & 3.4 & 1.7 & $0.20\pm0.02$ \\ 
216435.28 & CH$_3$CHO & $16_{2,15}$--$16_{1,16}$,~A & 2.7 & 1.2 & $0.06\pm0.01$ \\ 
216534.36 & CH$_3$CHO & $14_{3,11}$--$14_{2,12}$,~E & 2.7 & 1.8 & $0.10\pm0.02$ \\ 
216581.93 & CH$_3$CHO & $11_{1,10}$--$10_{1,9}$,~E & 3.2 & 2.7 & $0.15\pm0.02$ \\ 
216630.23 & CH$_3$CHO & $11_{1,10}$--$10_{1,9}$,~A & 3.3 & 3.1 & $0.22\pm0.02$ \\ 
216643.30 & SO$_2$ & $22_{2,20}$--$22_{1,21}$ & 2.2 & 4.5 & $0.50\pm0.04$ \\ 
216710.44 & H$_2$S & $2_{20}$--$2_{11}$ & 3.4 & 4.6 & $3.08\pm0.04$ \\ 
216830.15 & CH$_3$OCHO & $18_{2,16}$--$17_{2,15}$,~E & 3.3 & 6.6 & $0.27\pm0.03$ \\ 
216945.56 & CH$_3$OH & $5_{14}$--$4_{22}$  & 2.6 & 2.8 & $0.41\pm0.03$ \\ 
216967.42 & CH$_3$OCHO & $20_{0,20}$--$19_{0,19}$,~A & 4.4 & 9.9 & $0.92\pm0.04$ \\ 
217191.42 & CH$_3$OCH$_3$ & -- & 2.7 & 6.0 & $0.39\pm0.04$ \\ 
217238.54 & DCN & $J=3$--$2$ & 3.2 & 1.3 & $1.15\pm0.02$ \\ 
217822.15 & $c$-C$_3$H$_2$ & $6_{06}$--$5_{15}$, $6_{16}$--$5_{05}$ & 3.1 & 1.4 & $0.42\pm0.02$ \\ 
217886.39 & CH$_3$OH & $20_1$--$20_0$,~E1 & -0.0 & 2.7 & $0.13\pm0.03$ \\ 
217940.05 & $c$-C$_3$H$_2$ & $5_{14}$--$4_{23}$ & 3.6 & 1.9 & $0.41\pm0.02$ \\ 
218160.46 & $c$-C$_3$H$_2$ & $5_{24}$--$4_{13}$ & 2.6 & 0.8 & $0.07\pm0.01$ \\ 
218222.19 & H$_2$CO & $3_{03}$--$2_{02}$ & 3.2 & 1.1 & $5.05\pm0.02$ \\ 
218297.89 & CH$_3$OCHO & $17_{3,14}$--$16_{3,13}$,~A & 1.8 & 2.2 & $0.13\pm0.03$ \\ 
218324.72 & HC$_3$N & $J=24$--$23$ & 0.8 & 0.9 & $0.07\pm0.01$ \\ 
218440.06 & CH$_3$OH & $4_{2}$--$3_{1}$,~E  & 0.3 & 2.0 & $0.19\pm0.02$ \\ 
218475.63 & H$_2$CO & $3_{22}$--$2_{21}$ & 3.2 & 1.7 & $1.04\pm0.02$ \\ 
218654.66 & CH$_3$OCHO & $18_{16,2}$--$17_{16,1}$,~E & 3.3 & 2.9 & $0.10\pm0.02$ \\ 
218732.73 & $c$-C$_3$H$_2$ & $7_{16}$--$7_{07}$, $7_{26}$--$7_{17}$ & 2.4 & 1.7 & $0.02\pm0.02$ \\ 
218760.07 & H$_2$CO & $3_{21}$--$2_{20}$ & 3.5 & 2.8 & $2.33\pm0.02$ \\ 
218903.36 & OCS & $18$--$17$ & 2.5 & 1.1 & $0.24\pm0.02$ \\ 
218981.02 & HNCO & $10_{1,10}-9_{1,9}$ & 3.8 & 7.9 & $0.57\pm0.04$ \\ 
219355.01 & $^{34}$SO$_2$ & $11_{1,11}$--$10_{0,10}$ & 1.3 & 1.9 & $0.06\pm0.02$ \\ 
219560.35 & C$^{18}$O & $J=2$--$1$ & 3.1 & 1.3 & $7.70\pm0.02$ \\ 
219656.80 & HNCO & $10_{38}$--$9_{37}$, $10_{37}$--$9_{36}$ & 2.8 & 2.6 & $0.11\pm0.02$ \\ 
219737.19 & HNCO & $10_{28}$--$9_{27}$ & 6.3 & 10.1 & $0.64\pm0.04$ \\ 
219798.27 & HNCO & $10_{0,10}$--$9_{0,9}$ & 3.9 & 5.1 & $1.12\pm0.03$ \\ 
219820.39 & CH$_3$CHO & $4_{-2,3}$--$3_{-1,3}$,~E & 2.8 & 2.1 & $0.10\pm0.02$ \\ 
219908.52 & H$_2^{13}$CO & $3_{12}$--$2_{11}$ & 3.3 & 2.6 & $0.58\pm0.02$ \\ 
219949.44 & SO & $6_5$--$5_4$ & 3.6 & 1.5 & $4.79\pm0.02$ \\ 
\vspace{-6pt}\\
\multicolumn6l{\textbf{ISO-Oph 200}} \\ *
216112.58 & DCO$^+$ & $J=3$--$2$ & 4.7 & 0.5 & $0.26\pm0.03$ \\ 
218222.19 & H$_2$CO & $3_{03}$--$2_{02}$ & 4.7 & 0.8 & $0.70\pm0.03$ \\ 
219560.35 & C$^{18}$O & $J=2$--$1$ & 4.7 & 0.9 & $4.55\pm0.04$ \\ 
219949.44 & SO & $6_5$--$5_4$ & 4.7 & 0.9 & $0.48\pm0.03$ \\ 
\vspace{-6pt}\\
\multicolumn6l{\textbf{ISO-Oph 202}} \\ *
218222.19 & H$_2$CO & $3_{03}$--$2_{02}$ & 4.6 & 0.8 & $0.55\pm0.04$ \\ 
219560.35 & C$^{18}$O & $J=2$--$1$ & 4.6 & 0.7 & $3.59\pm0.04$ \\ 
219949.44 & SO & $6_5$--$5_4$ & 4.8 & 0.5 & $0.15\pm0.03$ \\ 
\vspace{-6pt}\\
\multicolumn6l{\textbf{ISO-Oph 203 (main)}} \\ *
216112.58 & DCO$^+$ & $J=3$--$2$ & 4.4 & 0.4 & $0.45\pm0.01$ \\ 
216278.76 & $c$-C$_3$H$_2$ & $3_{03}$--$2_{21}$ & 4.5 & 0.4 & $0.19\pm0.01$ \\ 
216373.32 & C$_2$D & $N=3$--$2,J=7/2$--$5/2$\tablenotemark{a} & 4.5 & 1.0 & $0.12\pm0.02$ \\ 
216428.32 & C$_2$D & $N=3$--$2,J=5/2$--$3/2$\tablenotemark{b} & 4.3 & 0.4 & $0.05\pm0.01$ \\ 
217822.15 & $c$-C$_3$H$_2$ & $6_{06}$--$5_{15}$, $6_{16}$--$5_{05}$ & 4.4 & 0.5 & $0.14\pm0.01$ \\ 
217940.05 & $c$-C$_3$H$_2$ & $5_{14}$--$4_{23}$ & 4.4 & 0.4 & $0.10\pm0.01$ \\ 
218222.19 & H$_2$CO & $3_{03}$--$2_{02}$ & 4.5 & 0.6 & $0.61\pm0.01$ \\ 
218440.06 & CH$_3$OH & $4_{2}$--$3_{1}$,~E  & 4.6 & 0.6 & $0.04\pm0.01$ \\ 
218475.63 & H$_2$CO & $3_{22}$--$2_{21}$ & 4.3 & 0.3 & $0.01\pm0.01$ \\ 
218760.07 & H$_2$CO & $3_{21}$--$2_{20}$ & 4.5 & 0.5 & $0.05\pm0.01$ \\ 
219560.35 & C$^{18}$O & $J=2$--$1$ & 4.5 & 0.7 & $5.63\pm0.02$ \\ 
219949.44 & SO & $6_5$--$5_4$ & 4.3 & 0.4 & $0.04\pm0.01$ \\ 
\vspace{-6pt}\\
\multicolumn6l{\textbf{ISO-Oph 203 (secondary)}} \\ *
216112.58 & DCO$^+$ & $J=3$--$2$ & 4.8 & 0.4 & $0.32\pm0.01$ \\ 
218222.19 & H$_2$CO & $3_{03}$--$2_{02}$ & 5.0 & 0.4 & $0.40\pm0.01$ \\ 
218440.06 & CH$_3$OH & $4_{2}$--$3_{1}$,~E  & 5.0 & 0.2 & $0.02\pm0.01$ \\ 
218475.63 & H$_2$CO & $3_{22}$--$2_{21}$ & 4.7 & 0.7 & $0.04\pm0.01$ \\ 
218760.07 & H$_2$CO & $3_{21}$--$2_{20}$ & 4.9 & 0.3 & $0.02\pm0.01$ \\ 
219560.35 & C$^{18}$O & $J=2$--$1$ & 4.9 & 0.3 & $0.76\pm0.01$ \\ 
219949.44 & SO & $6_5$--$5_4$ & 4.9 & 0.4 & $0.20\pm0.01$ \\ 
\vspace{-6pt}\\
\multicolumn6l{\textbf{ISO-Oph 209}} \\ *
216112.58 & DCO$^+$ & $J=3$--$2$ & 4.2 & 0.5 & $0.32\pm0.04$ \\ 
217822.15 & $c$-C$_3$H$_2$ & $6_{06}$--$5_{15}$, $6_{16}$--$5_{05}$ & 2.9 & 1.3 & $0.18\pm0.05$ \\ 
218222.19 & H$_2$CO & $3_{03}$--$2_{02}$ & 4.2 & 1.3 & $0.48\pm0.04$ \\ 
219560.35 & C$^{18}$O & $J=2$--$1$ & 4.4 & 1.3 & $3.42\pm0.06$ \\ 
219949.44 & SO & $6_5$--$5_4$ & 4.6 & 3.2 & $0.35\pm0.07$ \\ 
\vspace{-6pt}\\
\multicolumn6l{\textbf{GWAYL 4}} \\ *
216112.58 & DCO$^+$ & $J=3$--$2$ & 2.7 & 0.3 & $0.18\pm0.03$ \\ 
217822.15 & $c$-C$_3$H$_2$ & $6_{06}$--$5_{15}$, $6_{16}$--$5_{05}$ & 3.0 & 1.3 & $0.14\pm0.04$ \\ 
218222.19 & H$_2$CO & $3_{03}$--$2_{02}$ & 2.8 & 0.5 & $0.41\pm0.03$ \\ 
219560.35 & C$^{18}$O & $J=2$--$1$ & 2.7 & 0.6 & $4.76\pm0.04$ \\ 
219949.44 & SO & $6_5$--$5_4$ & 3.0 & 1.8 & $0.54\pm0.05$ \\ 
\vspace{-6pt}\\
\multicolumn6l{\textbf{Oph-emb 4}} \\ *
219560.35 & C$^{18}$O & $J=2$--$1$ & 2.6 & 0.5 & $3.21\pm0.04$ \\ 
\vspace{-6pt}\\
\multicolumn6l{\textbf{Elias 28}} \\ *
219560.35 & C$^{18}$O & $J=2$--$1$ & 4.2 & 1.0 & $1.27\pm0.06$ \\ 
\vspace{-6pt}\\
\multicolumn6l{\textbf{J162624}} \\ *
218222.19 & H$_2$CO & $3_{03}$--$2_{02}$ & 3.3 & 0.8 & $0.33\pm0.03$ \\ 
219560.35 & C$^{18}$O & $J=2$--$1$ & ... & ... & Off-em. \\ 
219949.44 & SO & $6_5$--$5_4$ & 3.5 & 0.9 & $0.41\pm0.04$ \\ 
\vspace{-6pt}\\
\multicolumn6l{\textbf{J1633.92442}} \\ *
\multicolumn6l{No lines detected} \\ 
\vspace{-6pt}\\
\multicolumn6l{\textbf{MMS126}} \\ *
216112.58 & DCO$^+$ & $J=3$--$2$ & 4.0 & 0.5 & $1.16\pm0.02$ \\ 
216278.76 & $c$-C$_3$H$_2$ & $3_{03}$--$2_{21}$ & 4.0 & 0.5 & $0.42\pm0.02$ \\ 
216373.32 & C$_2$D & $N=3$--$2,J=7/2$--$5/2$\tablenotemark{a} & 4.2 & 1.0 & $0.08\pm0.02$ \\ 
217238.54 & DCN & $J=3$--$2$ & 4.2 & 0.6 & $0.06\pm0.02$ \\ 
217822.15 & $c$-C$_3$H$_2$ & $6_{06}$--$5_{15}$, $6_{16}$--$5_{05}$ & 4.1 & 0.4 & $0.28\pm0.02$ \\ 
217940.05 & $c$-C$_3$H$_2$ & $5_{14}$--$4_{23}$ & 4.1 & 0.3 & $0.19\pm0.01$ \\ 
218160.46 & $c$-C$_3$H$_2$ & $5_{24}$--$4_{13}$ & 4.1 & 0.2 & $0.05\pm0.01$ \\ 
218222.19 & H$_2$CO & $3_{03}$--$2_{02}$ & 4.1 & 0.4 & $0.88\pm0.02$ \\ 
218440.06 & CH$_3$OH & $4_{2}$--$3_{1}$,~E  & 3.6 & 1.0 & $0.08\pm0.02$ \\ 
219560.35 & C$^{18}$O & $J=2$--$1$ & 4.1 & 0.9 & $4.45\pm0.02$ \\ 
219908.52 & H$_2^{13}$CO & $3_{12}$--$2_{11}$ & 4.4 & 2.6 & $0.13\pm0.03$ \\ 
219949.44 & SO & $6_5$--$5_4$ & 4.2 & 0.5 & $0.23\pm0.02$ \\ 
\vspace{-6pt}\\
\multicolumn6l{\textbf{Oph-emb 18}} \\ *
218222.19 & H$_2$CO & $3_{03}$--$2_{02}$ & 3.5 & 0.2 & $0.06\pm0.02$ \\ 
218440.06 & CH$_3$OH & $4_{2}$--$3_{1}$,~E  & 2.0 & 3.8 & $0.18\pm0.05$ \\ 
219560.35 & C$^{18}$O & $J=2$--$1$ & 3.5 & 0.5 & $3.97\pm0.03$ 
	
	\enddata
	
	\tablenotetext{a}{Blend of $F=9/2$--$7/2$, $F=7/2$--$5/2$, and $F=5/2$--$3/2$.}
	\tablenotetext{b}{Blend of $F=7/2$--$5/2$, $F=5/2$--$3/2$, and $F=3/2$--$1/2$.}
	\tablenotetext{c}{Blend of $F=13/2$--$11/2$ and $F=15/2$--$13/2$.}
	\tablenotetext{d}{$F_1=2$--$2, F=2$--$1$.}
	\tablenotetext{e}{$F_1=2$--$1, F=2$--$1$.}
	\tablenotetext{f}{$F_1=2$--$1, F=3$--$2$.}
	\tablenotetext{g}{$F_1=3$--$2, F=2$--$1$.}
	%% Include any \tablenotetext{key}{text}, \tablerefs{ref list},
	%% or \tablecomments{text} between the \enddata and 
	%% \end{deluxetable} commands
	
	%% No \tablecomments indicated
	
	%% No \tablerefs indicated
	\tablecomments{Rest frequencies from the CDMS \citep{cdms} and JPL \citep{jpl} molecular spectroscopy databases.}
	
	\label{tab:obslines}
\end{deluxetable*}
\capstarttrue
	
\section{Spectral line profiles in Corona Australis sources}
\label{sec:cra}

\begin{figure*}[b]
	\epsscale{1.2}
	\plotone{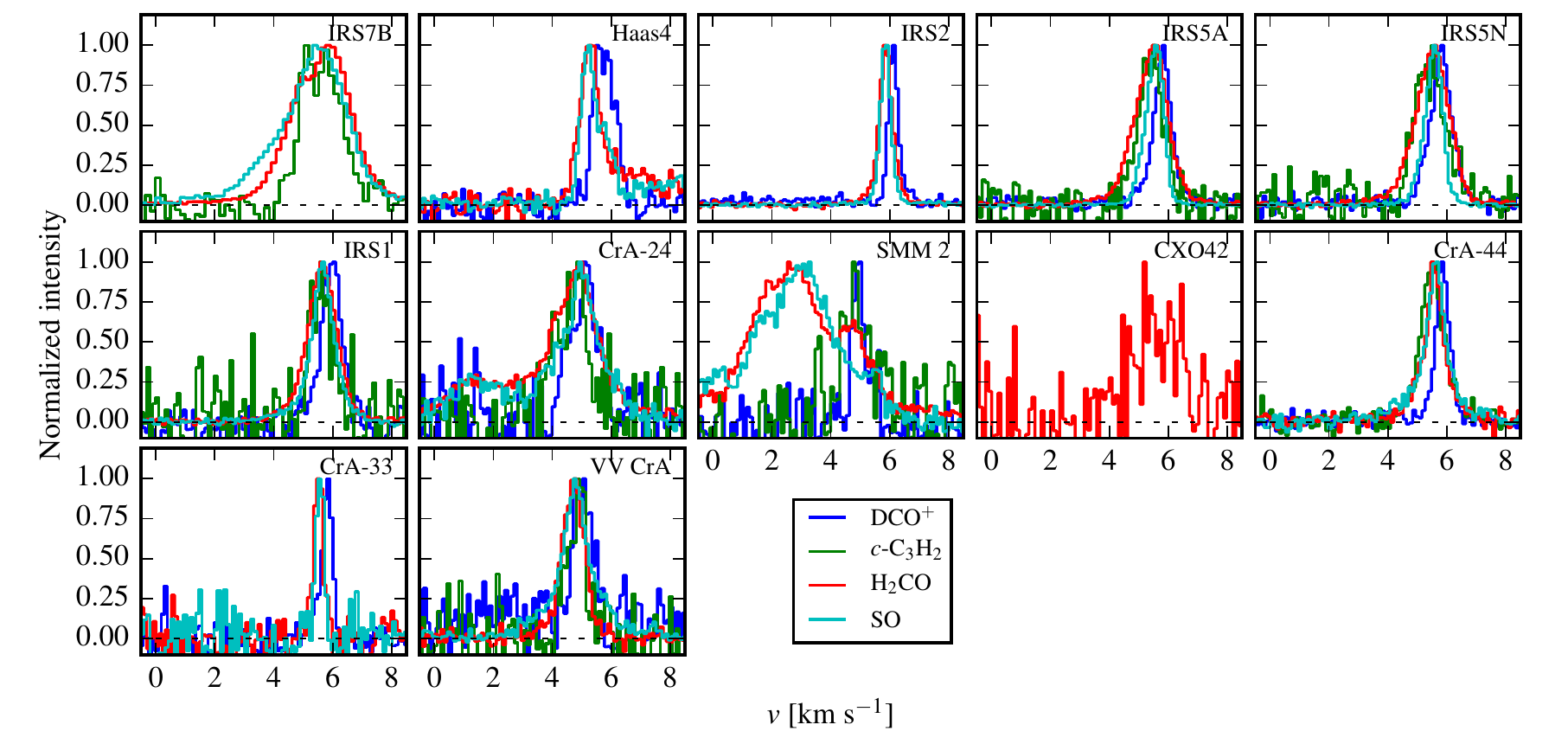}
	\caption{Normalized spectra of the DCO$^+$ (216.113~GHz), \ctht\ (217.822~GHz), H$_2$CO (218.222~GHz), and SO (219.949~GHz) lines for all sources in the CrA survey of embedded protostars \citep{lindberg15} for which at least one of these lines is detected at a $3\sigma$ level or higher (CrA-3, CrA-5, LS-RCrA1, CrA-37, and CrA-46 are thus not included). The DCO$^+$ line was not covered by the R~CrA IRS7B observations.}
	\label{fig:allspectra_cra}
\end{figure*}

For completeness, \autoref{fig:allspectra_cra} shows the spectral profiles of the sources in the CrA survey of embedded protostars \citep{lindberg15} by plotting the normalized spectra for the strongest DCO$^+$ (216.113~GHz), \ctht\ (217.822~GHz), H$_2$CO (218.222~GHz), and SO (219.949~GHz) lines in each source where they have been detected (cf. \autoref{allspectra}).

\section{Non-LTE analysis of \textit{\lowercase{c}}\nobr C$_3$H$_2$}
\label{sec:c3h2}

In our observations, the rotational temperatures of \ctht\ were generally found to be lower than those of H$_2$CO. To verify that this reflects a difference in physical temperatures and is not just a non-LTE or optical depth effect, we used the radiative transfer code RADEX \citep{radex} to study line ratios within a range of physical parameters. A similar investigation of the H$_2$CO rotational temperature was performed by \citet{mangum93}, showing that a rotational temperature calculated from H$_2$CO transitions with the same $J_\mathrm{u}$ will provide a good measure of the physical temperature.

We here use RADEX to calculate non-LTE line ratios of four of the five \ctht\ lines covered in our observations (see \autoref{tab:spectrallines}; the 218.733~GHz line is not included since it is only detected toward one source). One of these lines is a blend between an ortho and a para line, two lines are ortho lines, and the final one is a para line. Excitation analysis involving a combination of ortho and para lines will therefore depend on the assumed ortho-to-para ratio. For \ctht, we use a ratio of 3 \citep{lucas00}. We use the collisional rates of \citet{chandra00} retrieved from the LAMDA database \citep{lamda}.

\begin{figure*}
	\epsscale{1.1}
	\plottwo{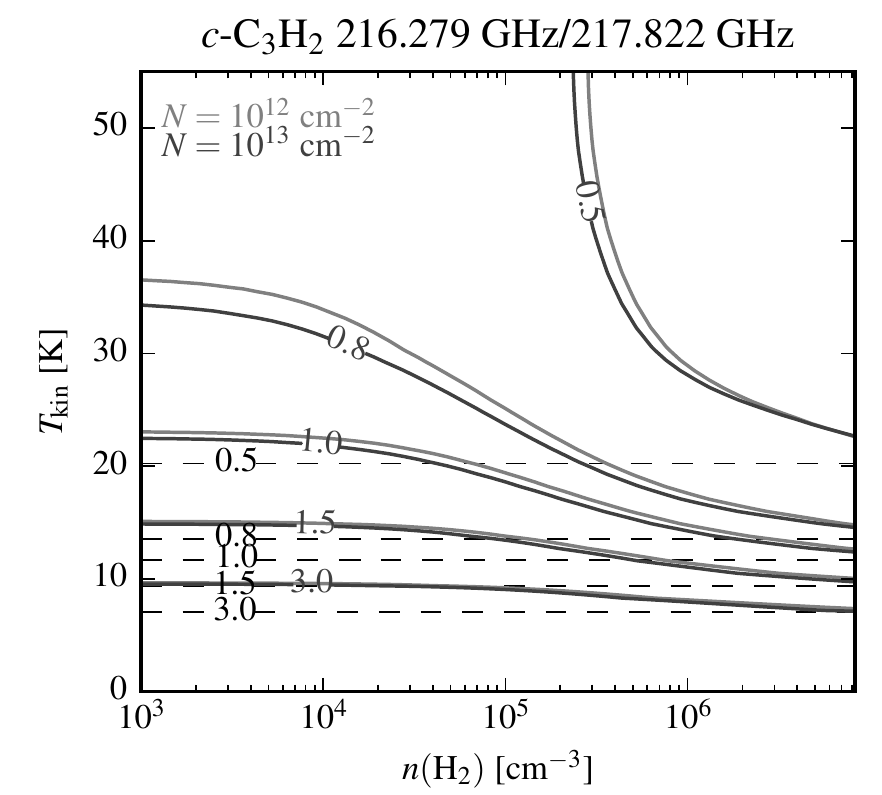}{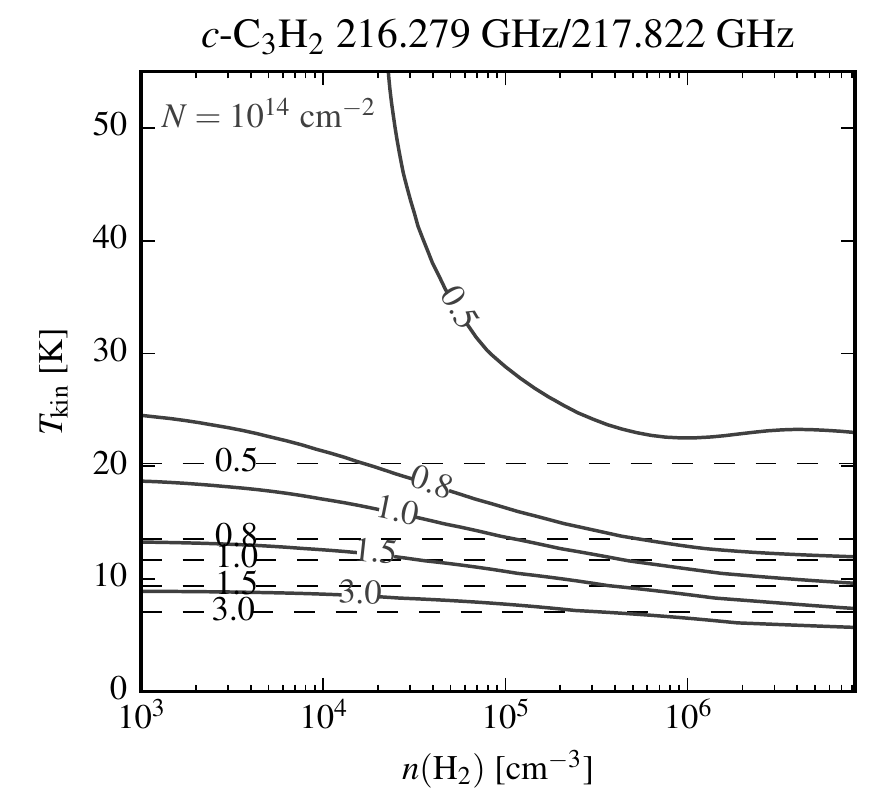}
	\plottwo{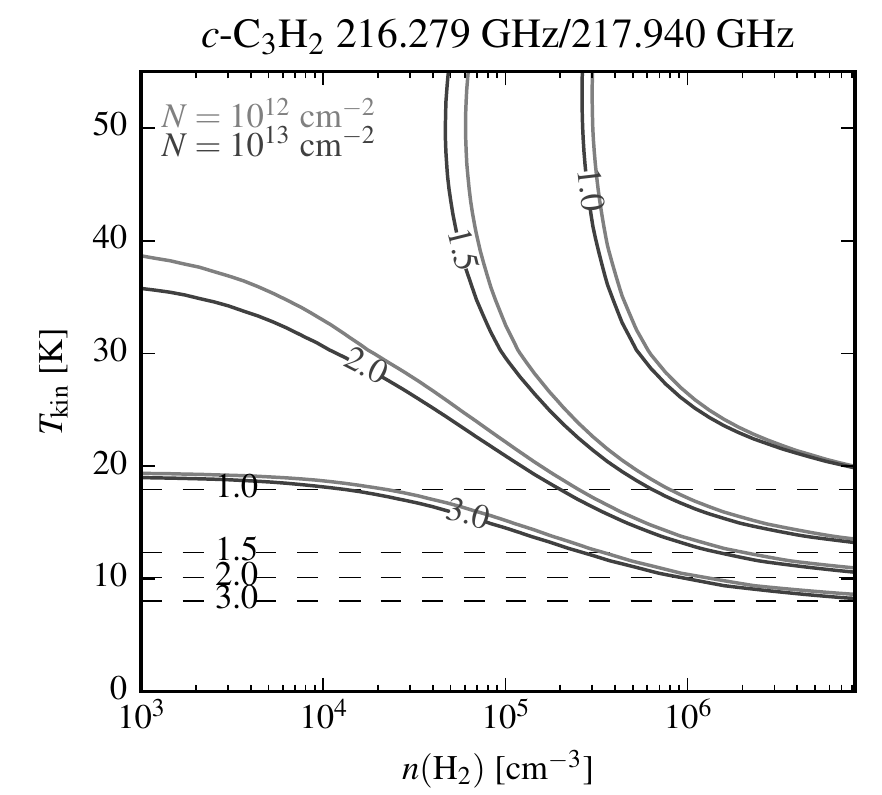}{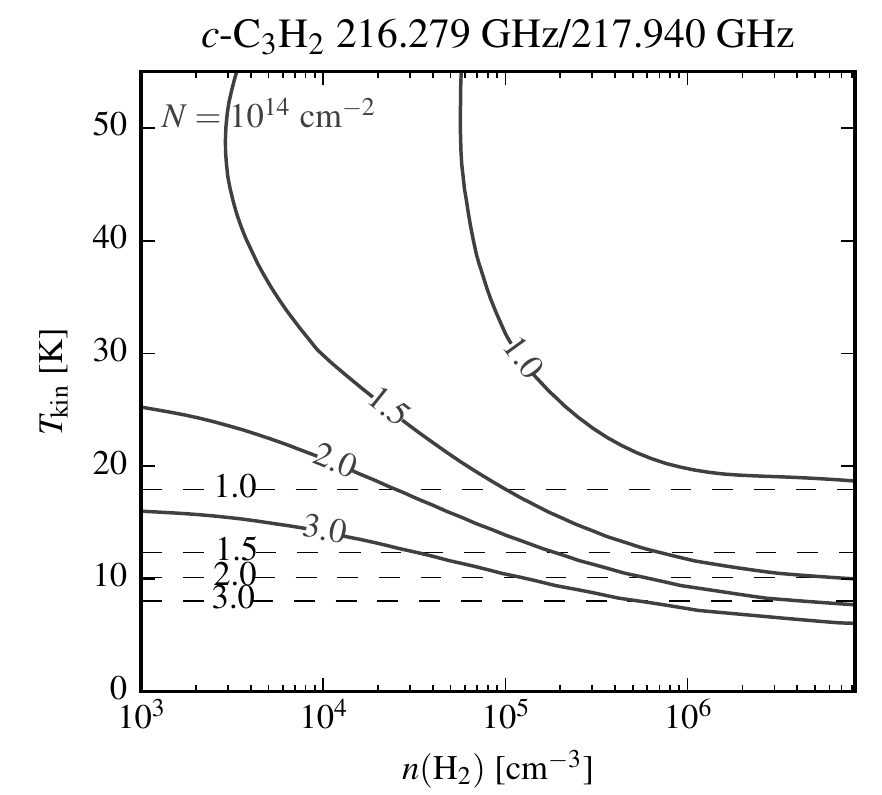}
	\plottwo{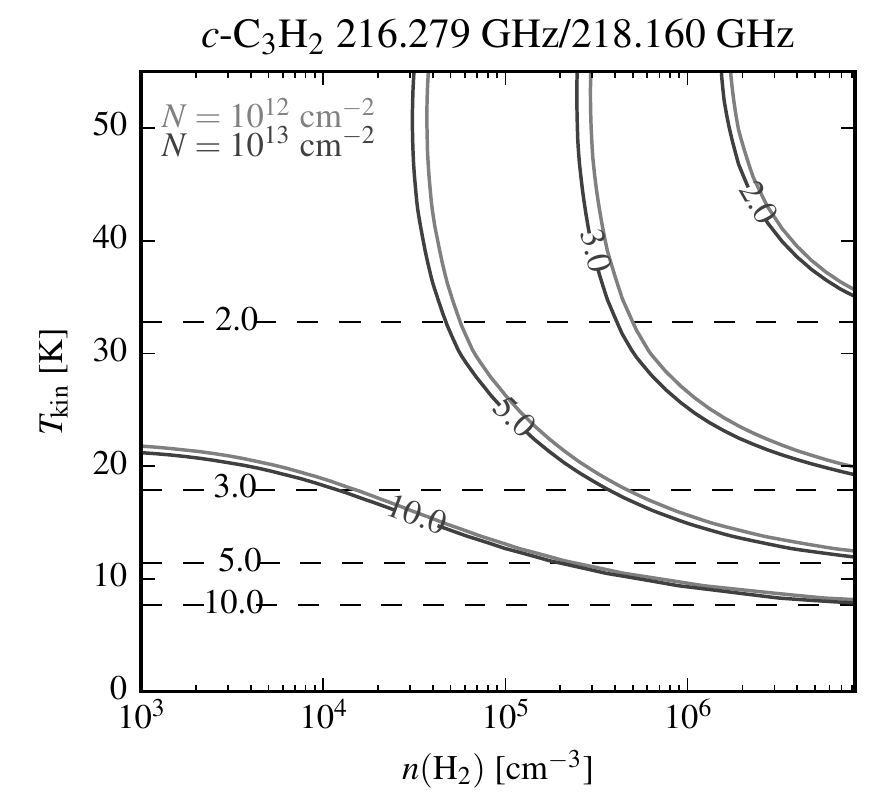}{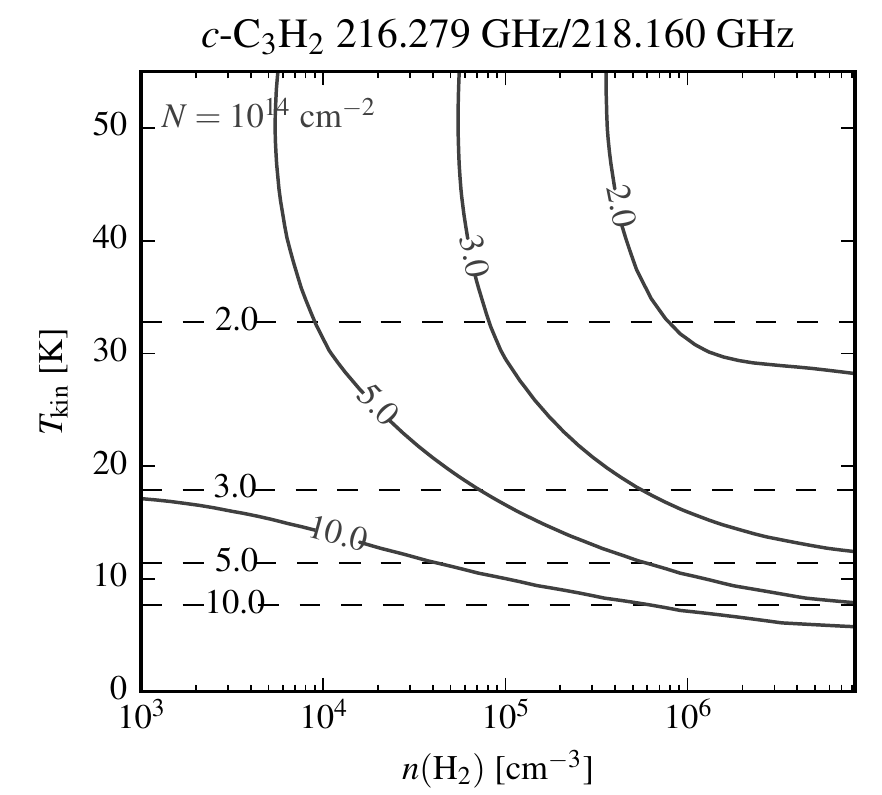}
	\caption{\ctht\ line ratios assuming LTE (dashed lines) and with non-LTE RADEX models (solid lines).}
	\label{fig:c3h2}
\end{figure*}

Since three of the four lines have roughly the same upper level energy ($E_\mathrm{u} =35$--39~K), we only show ratios between the 216.279~GHz line ($E_\mathrm{u} = 19.5$~K) and the remaining three lines. We perform the non-LTE calculations assuming three different column densities (total ortho+para \ctht\ column densities): $N=10^{12}$~cm$^{-2}$, $N=10^{13}$~cm$^{-2}$, and $N=10^{14}$~cm$^{-2}$. Our observations show that the \ctht\ column densities typically are $<10^{13}$~cm$^{-2}$, so the $N=10^{14}$~cm$^{-2}$ case is unlikely to be applicable to this work. A model with $N=10^{11}$~cm$^{-2}$ (not plotted) has results identical to the $N=10^{12}$~cm$^{-2}$ model. The results are shown in \autoref{fig:c3h2}.

We find that the 216.279~GHz/217.822~GHz non-LTE line ratio is within $\sim 5$~K of the LTE solution given $T_\mathrm{kin}\lesssim20$~K for all densities and column densities. The other two ratios are worse off in predicting the temperature, in particular for low H$_2$ densities. At $T_\mathrm{kin}\lesssim20$~K and $n(\mathrm{H}_2)\gtrsim5\times10^5$~cm$^{-3}$, the temperature is, however, only under-predicted (or over-predicted) by $\lesssim5$~K. Densities are expected to exceed $5\times10^5$~cm$^{-3}$ within the inner 1800~au of a majority of embedded protostars \citep[e.g.][]{jorgensen02}, which is the radius of the APEX beam projected to the distance of the Ophiuchus cloud (125~pc).

We compare the rotational (LTE) temperatures calculated when using only the 216.279~GHz and the 217.822~GHz lines with those calculated from using all detected lines for all sources in the sample, and find that in no source the difference is greater than 2~K. The rotational temperatures of \ctht\ presented in \autoref{tab:rotdiagresults} use all observed \ctht\ lines, and might thus be under-estimated, but calculating rotational temperatures from RADEX intensities shows that this difference is $\lesssim5$~K. We therefore conclude that the physical temperature of the  \ctht\ gas is significantly lower than that of the H$_2$CO gas in the externally irradiated sources in the sample.

\end{appendix}

\end{document}